\documentclass[pra,a4paper,twocolumn,showpacs,superscriptaddress]{revtex4}
\usepackage{amsmath,amscd,amsfonts,amssymb, amsbsy, color,bbm, bm}
\usepackage{graphicx}
\usepackage{enumitem}
\newcommand{\be}{\begin{equation}}
\newcommand{\ee}{\end{equation}}

\newcommand{\ba}[1]{\left(\begin{array}{#1}}
\newcommand{\ea}{\end{array}\right)}

\numberwithin{equation}{section}
\renewcommand{\theequation}{\arabic{section}.\arabic{equation}}
\begin{document}
\title{Canonical forms of two-qubit states under local operations} 
\author{Sudha} 
\affiliation{Department of Physics, Kuvempu University, 
	Shankaraghatta-577 451, Karnataka, India}
\affiliation{Inspire Institute Inc., Alexandria, Virginia, 22303, USA.}
\author{H. S. Karthik} 
\affiliation{International Centre for Theory of Quantum Technologies, University of Gdansk, Gdansk, Poland}
\author{Rajarshi Pal}\affiliation{Department of Physics,
	Sungkyunkwan University, Suwon 16419, Korea}
	\author{K. S. Akhilesh} 
\affiliation{Department of Studies in Physics, University of Mysore, Manasagangotri, Mysuru-570006, Karnataka, India}
\author{Sibasish Ghosh}
\affiliation{Optics $\&$ Quantum Information Group, The Institute of Mathematical Sciences, HBNI,	C. I. T. Campus, Taramani, Chennai 600113, India}
\author{K. S. Mallesh} 
\affiliation{Department of Studies in Physics, University of Mysore, Manasagangotri, Mysuru 570006, Karnataka, India}
\affiliation{Regional Institute of Education (NCERT), Mysuru 570006, India}
\author{A. R. Usha Devi} 
\affiliation{Department of Physics, Bangalore University, 
	Bangalore-560 056, India}
\affiliation{Inspire Institute Inc., Alexandria, Virginia, 22303, USA.}
\email{arutth@rediffmail.com} 

\date{\today}
\begin{abstract} 
Canonical forms of two-qubits under the action of stochastic local operations and classical communications (SLOCC) offer great insight for understanding non-locality and entanglement shared by them. They also enable  geometric picture of two-qubit states within the Bloch ball. It has been shown (Verstraete et.al. (\pra {\bf 64}, 010101(R) (2001)) that an arbitrary two-qubit state gets transformed under SLOCC into one of the {\em two} different canonical forms. One of these happens to be the Bell diagonal form of two-qubit states and the other non-diagonal canonical form is obtained for a family of rank deficient two-qubit states. The method employed by Verstraete et.al. required   highly non-trivial results on matrix decompositions in $n$ dimensional spaces with indefinite metric.  Here we employ an entirely different approach -- inspired by the methods developed by Rao et. al., (J. Mod. Opt. {\bf 45}, 955 (1998)) in classical polarization optics --  which leads naturally towards the identification of  two inequivalent SLOCC invariant canonical forms for two-qubit states.  In addition, our approach results in  a simple geometric visualization of  two-qubit states in terms of their SLOCC canonical forms. 
\end{abstract}
\pacs{03.65.Ta, 03.67.Mn}
\maketitle
\section{Introduction} 
Geometric intuition inscribed in the Bloch ball picture of qubits serves as  a  powerful tool in the field of quantum information processing. Simplicity of this geometric  representation of qubits inspired its generalization to quantum single party systems in higher dimensions~\cite{Kimura2003, Kimura2005, Sandeep2016}. However, these attempts resulted in complicated geometric features. On the other hand, physically relevant visualization of the simplest bipartite quantum system viz., joint state of two-qubits, has been developed by several groups~\cite{hor96, verstraete2001, verstraete2002, zyzcBook2006,avron2007,avron2009,shi2011,shi2012,jevtic2014,gamel2016}.  Geometric representation of two-qubit states inside the Bloch ball provides a natural picture to understand correlation properties like entanglement~\cite{hor96, verstraete2002, jevtic2014, MilneNJP2014, MilnePRA2014, gamel2016}, quantum discord~\cite{shi2011,shi2012,jevtic2014},  and non-local steering~\cite{verstraete2002,jevtic2014,jevtic2015, MilneNJP2014,MilnePRA2014,Ngu2016,NguEPL2016,Quan2016}.  

Previously, Verstraete et.al.~\cite{verstraete2001,verstraete2002} highlighted that SLOCC on a two-qubit density matrix $\rho_{AB}$ correspond to Lorentz transformations on the $4\times 4$ real matrix parametrization $\Lambda$ of $\rho_{AB}$ and they arrived at two different types of canonical forms for the real matrix $\Lambda$. The canonical forms of $\Lambda$ correspond to its Lorentz singular value decompositions, offering a natural classification of the set of all two-qubit density matrices into {\em two} different SLOCC families. The canonical SLOCC transformations also paved way  to visualization of  two-qubit state --  as an ellipsoid inscribed inside the Bloch ball~\cite{verstraete2001,verstraete2002, shi2011, shi2012,jevtic2014}. However, 
the mathematical recipe used  in   Ref.~\cite{verstraete2001,verstraete2002} to arrive at the  SLOCC canonical forms is highly technical and depended on non-trivial results on  matrix decompositions in spaces with indefinite metric~\cite{Gohberg83}. Moreover, it was pointed out~\cite{jevtic2014} that this approach fails to reveal the geometric features in an unambiguous fashion. A more detailed investigation by Jevtic et.al.,~\cite{jevtic2014} focussed towards an elegant  geometric representation, mapping a two-qubit state to an ellipsoid lying inside the Bloch ball, in a complete manner with the help of suitable SLOCC transformations. However, this work did not address the relevant issue of identifying canonical forms of two-qubit density matrix $\rho_{AB}$, based on the Lorentz singular value decomposition of the associated  $4\times 4$ real matrix $\Lambda$.  A straightforward method to identify  Lorentz singular value decomposition, which in turn gets connected with the SLOCC canonical forms of two-qubit states, is still lacking. In this paper we address this issue, using the methods developed in classical polarization optics by some of us~\cite{AVG1,AVG2}. Our method leads to the identification of two different  types of SLOCC canonical forms for two-qubit states. The canonical forms identified by our approach are shown to be Lorentz equivalent to the ones obtained in Ref.~\cite{verstraete2001}. Our detailed analysis  gives a fresh perspective on the geometric representation of two-qubit states in terms of their SLOCC inequivalent canonical forms. 
 
Contents of the paper are organized as follows: In Sec.~2, we obtain the $4\times 4$ real matrix parametrization $\Lambda$ of a two-qubit density matrix  $\rho_{AB}$ shared between  Alice and Bob. After giving a 
 brief outline on Minkowski space notions of  positive, neutral, negative four-vectors and orthochronous proper Lorentz group (OPLG), we show that 
 (i)  the real $4\times 4$ matrix $\Lambda$  gets pre and post multiplied by the $4\times 4$ Lorentz matrices $L_A$ and $L_B^T$ (the superscript ``$T$'' denotes matrix transposition) under the action of SLOCC transformation -- implemented by Alice and Bob respectively on their individual  qubits; (ii)  the $4\times 4$ real matrix  $\Lambda$ maps the set of all four-vectors with non-negative Minkowski norm  into itself. Sec.~3 gives details on finding the canonical forms of real symmetric matrices $\Omega_{A}=\Lambda\, G\, \Lambda^T$  and $\Omega_{B}=\Lambda^T\, G\, \Lambda$, (which are constructed from the real matrix $\Lambda$ and  the Minkowski metric $G={\rm diag}\,(1,-1,-1,-1)$) using  Lorentz congruent transformations $L_A\, \Omega_{A}\, L_A^T$ and $L_B\, \Omega_{B}\, L_B^T$ respectively. Lorentz singular value decompositions $\Lambda^c=L_A\, \Lambda\, L_B^T$ of two inequivalent canonical forms  $ \Lambda^{\rm{I}_c}$,  $\Lambda^{\rm{II}_c}$ of  the  $4\times 4$ real matrix $\Lambda$ and the corresponding two-qubit density matrices $ \rho_{AB}^{\rm{I}_c}$,  $\rho_{AB}^{\rm{II}_c}, \rho^{\rm{II}_c}_{BA} $ are also given here. Furthermore, equivalence between the canonical forms obtained earlier by Verstraete et. al.~\cite{verstraete2001,verstraete2002} with the ones realized based on our approach, is established in Sec.~3. Geometric representation to aid visualization of SLOCC  canonical forms of the two-qubit states is discussed in Sec.~4. A consise summary of our results is presented in  Sec.~5.  

\section{Real parametrization of two-qubit density matrix and SLOCC transformations} 
Consider a  two-qubit state $\rho_{AB}$ belonging to the Hilbert space ${\mathcal{H}}_A\otimes {\mathcal{H}}_B\equiv{\mathbbm{C}}_2\otimes {\mathbbm{C}}_2$, shared between two parties Alice and Bob. It can be expressed in the Hilbert-Schmidt basis $\{\sigma_\mu\otimes \sigma_\nu, \mu,\nu=0,1,2,3\}$ as,   
\begin{eqnarray}
\label{rho}
\rho_{AB}&=&\frac{1}{4}\, \sum_{\mu,\,\nu=0}^{3}\,   
\Lambda_{\mu \, \nu}\, \left( \sigma_\mu\otimes\sigma_\nu \right)  
\end{eqnarray}
where 
\begin{eqnarray}
\label{lambda}
\Lambda_{\mu \, \nu}&=& {\rm Tr}\,\left[\rho_{AB}\,
(\sigma_\mu\otimes\sigma_\nu)\,\right].   
\end{eqnarray} 
Here $\sigma_0=\mathbbm{1}_2$, denotes $2\times 2$ identity matrix and $\sigma_1,\sigma_2,\sigma_3$ are the Pauli matrices.

Expressed in the $2\times 2$ block form, the $4\times 4$ real matrix $\Lambda$  defined in  (\ref{lambda})  takes the following compact form,   
\begin{eqnarray}
\label{block}
\Lambda&=&\left(\begin{array}{ll} 1& {\bf b}^T  \\ {\bf a}  &  T   
\end{array}\right),
\end{eqnarray} 
with the superscript ``$T\,$'' denoting matrix transposition; ${\bf a}=(a_1,\, a_2,\, a_3)^T$, ${\bf b}=(b_1,\, b_2,\ b_3)^T$ denote  Bloch vectors  of the reduced density matrices $\rho_A={\rm Tr}_B(\rho_{AB})$, $\rho_B={\rm Tr}_A(\rho_{AB})$ of qubits $A$, $B$ respectively and $T$ corresponds to $3\times 3$ real correlation matrix, elements of which are given by $t_{ij}={\rm Tr}(\rho_{AB}\,\sigma_i\otimes\sigma_j),\, i,j=1,2,3$. Thus, the $4\times 4$  matrix $\Lambda$ is characterized by 15 real parameters (3 each of the   Bloch vectors ${\bf a},\ {\bf b}$ and 9 elements of the correlation matrix $T$) and provides a unique {\em real matrix parametrization} of the two-qubit density matrix 
$\rho_{AB}$. 

We give a brief outline on the Minkowski four-vectors and OPLG transformations in the following subsection, before discussing the properties of the real parametrization $\Lambda$ of the two-qubit state.
 

\subsection{Minkowski space, four-vectors, and OPLG}\label{2a}
The Minkowski space ${\mathcal{M}}$ is a  four dimensional real vector space consisting of four-vectors or Minkowski vectors~\cite{synge, KNS},  denoted by $\mathbf{x}=(x_0,\, x_1, x_2, x_3)^T$. The space is equipped with the metric 
\begin{equation}
 \label{g} 
 G={\rm diag}\, (1,-1,-1,-1),
 \end{equation} 
and a scalar product 
\begin{equation} 
\mathbf{x}^T\, G\, \mathbf{y}=x_0\, y_0-x_1\, y_1-x_2\, y_2-x_3\, y_3.
\end{equation}
As the Minkowski squared norm  $\mathbf{x}^T\,G\,\mathbf{x}$ of an arbitrary four-vector $\mathbf{x}$ can assume positive, zero, or negative values, we employ the following nomenclature~\cite{AVG1,AVG2,KNS,synge,note}:  
\begin{eqnarray*}
\begin{array}{lll}
{\rm (i)} & \mathbf{x}^T\,G\,\mathbf{x}>0: \hskip 0.1in & \ {\rm positive\ four-vector} \\ 
{\rm (ii)} & \mathbf{x}^T\,G\,\mathbf{x}=0:  & \ {\rm neutral\ (or\ null)\ four-vector} \\ 
{\rm (iii)} & \mathbf{x}^T\,G\,\mathbf{x}<0:  & \ {\rm negative\ four-vector} 
\end{array}
\end{eqnarray*}

Consider the set of all real $4\times 4$ matrices    
$$\{L\mid \, \det\,L=1,\, L_{00}>0  \}, $$  
which constitutes the orthochronous proper Lorentz group  SO(3,1). By definition, Lorentz matrix $L$  preserves the Minkowski norm  i.e., the four-vector $L\, \mathbf{x}$ is positive, neutral or negative depending on whether $\mathbf{x}\in{\mathcal{M}}$ is   positive, neutral or negative respectively. In particular, it is pertinent to highlight that the set   
$S_+:\left\{\mathbf{x}\in{\mathcal{M}}\, \left\vert\, {\mathbf{x}^T}\, G\, \mathbf{x}\geq 0,\ x_0 > 0\right.\right\}$ of four-vectors gets mapped to itself under OPLG i.e., $\tilde{S}_+:\left\{\tilde{\mathbf{x}}=L\, \mathbf{x}\in{\mathcal{M}}\, \left\vert\, \tilde{\mathbf{x}}^T\, G\  \tilde{\mathbf{x}}\geq 0,\  \tilde{x}_0> 0\right.\right\}\equiv S_+$. We would discuss, in subsection 2C, about encoding the set of all  non-negative single qubit operators in $\mathbbm{C}_2$ in terms of four-vectors of the set  $S_+$.

\subsection{SLOCC transformations and OPLG}

Under the action of stochastic local operations and classical communication
(SLOCC), a two-qubit density matrix $\rho_{AB}$ transforms as~\cite{verstraete2001,verstraete2002,jevtic2014} 
\be
\label{slocc2}
\rho_{AB}\longrightarrow\, \widetilde{\rho}_{AB} =\frac{\left(A\otimes\, B\right)\, \rho_{AB}\, \left(A^\dag\,\otimes B^\dag\right)}{{\rm Tr}\left[\,\rho_{AB}\,(\, A^\dag\, A\otimes B^\dag\, B\, )\,\right]},
\ee
where $A, B\in {\rm SL(2,C)}$ denote $2\times 2$ complex matrices with unit determinant. Owing to the  homomorphism between the groups  SL(2,C) and   
 SO(3,1), one finds the correspondence  $\pm A \mapsto L_A,\, \pm B \mapsto L_B $, where $A,\,B\in$ SL(2,C) and $L_A,\, L_B\in$ SO(3,1). In particular, the basis matrices     $\sigma_\mu\otimes\sigma_\nu, \mu,\, \nu=0,1,2,3$  get transformed  under SL(2,C)\,$\otimes$\,SL(2,C) as   
\begin{eqnarray}
\label{oplt}
&&(A\otimes B)(\sigma_\mu\otimes\sigma_\nu)(A^\dag\otimes B^\dag) = 
A\,\sigma_\mu A^\dag
\otimes B\,\sigma_\nu B^\dag    \nonumber \\
&& \hskip 0.5in =  \sum_{\alpha,\beta=0,1,2,3} \left(L_{A}\right)_{\alpha\mu} \left(L_{B}\right)_{\beta\nu} \sigma_\alpha\otimes\sigma_\beta.  
\end{eqnarray} 
Thus, SLOCC operation  $\rho_{AB}~\rightarrow~ \tilde{\rho}_{AB}$ on the two-qubit state is equivalent to the following transformation (up to normalization)
\begin{eqnarray}
\label{sl2c}
\Lambda\longrightarrow \tilde{\Lambda}&=&\, L_A\,\Lambda\, L_B^{T}
\end{eqnarray}  
on the real matrix $\Lambda$. So, it is evident that  $\tilde{\Lambda}$ --  obtained after OPLG transformations $L_A, L_B$  on $\Lambda$  (see (\ref{sl2c})) --   parametrizes  the two-qubit density matrix $\tilde{\rho}_{AB}$, which is  physically realizable under SLOCC.  Using  suitable OPLG transformations $L_{A_c}$, $L_{B_c}$ one should be able to arrive at a  canonical (normal) form   $\Lambda^{c}$   associated with a given $\Lambda$  i.e., 
\begin{eqnarray}
\label{can}
\Lambda^c=\, L_{A_c}\,\Lambda\, L_{B_c}^{T}.
\end{eqnarray}  
It may be seen that (\ref{can}) is the Minkowski space counterpart of the  singular value decomposition in Euclidian space and  is referred to as the Lorentz singular value decomposition~\cite{verstraete2001,verstraete2002}.   

\subsection{Real symmetric matrices $\mathbf{\Omega_A=\Lambda\,G\,\Lambda^T}$ and $\mathbf{\Omega_B=\Lambda^T\,G\,\Lambda}$}
Let us denote the set  of all non-negative operators acting on the Hilbert space $\mathbbm{C}_2$ by ${\mathcal{P}}^+ :=\{P \vert P\, \geq\, 0\, \}$. An element  $P\in {\mathcal{P}}^+$ can be represented in the Pauli basis $\sigma_\mu=(\mathbbm{1}_2,\sigma_1,\sigma_2,\sigma_3)$ as 
\begin{equation}
\label{pmu}
P=\frac{1}{2}\,\sum_{\mu}\, p_\mu\,  \sigma_\mu
\end{equation}   
where $p_\mu= {\rm Tr}(P\, \sigma_\mu), \mu=0,1,2,3$ are the four real parameters characterizing $P$. With every $P\in {\mathcal{P}}^+$, we associate  a four-vector  $\mathbf{p}=(p_0,\, p_1,\, p_2,\, p_3)^T$. Non-negativity  $P\geq 0$ of the operator $P$  is synonymous to the conditions  $p_0> 0$ and $p_0^2-p_1^2-p_2^2-p_3^2\geq 0$ on the four-vector $\mathbf{p}$.  In the language of  Minkowski space,  non-negativity of the operator $P\geq 0$ reflects itself as the squared Minkowski norm condition  $\mathbf{p}^T\, G\, \mathbf{p}\geq 0$  together with the restriction $p_0> 0$ on zeroth component of the four-vector $\mathbf{p}$.   

Let us consider the map 
\begin{eqnarray}
\label{mapA2B}
P_A \mapsto Q_B&=&2\, {\rm Tr}_A\left[(\sqrt{P_A}\otimes \mathbbm{1}_2)\, \rho_{AB}\,(\sqrt{P_A}\otimes \mathbbm{1}_2)\right] \nonumber \\ 
&=& 2\, {\rm Tr}_A\left[\rho_{AB}\, (P_A\otimes \mathbbm{1}_2)\right]
\end{eqnarray}
from  the set of all non-negative operators ${\mathcal{P}}_A^+~:=~ \{P_A\, \vert\ P_A\geq 0 \}$ on the Hilbert space ${\mathcal{H}}_A$  to the set of non-negative operators ${\mathcal{Q}}^+_B:=\, \{Q_B\, =2\,  {\rm Tr}_A[\rho_{AB}\,(P_A\otimes \mathbbm{1}_2)]\}$ acting on the Hilbert space ${\mathcal{H}}_B$. We have,  
\begin{eqnarray}
\label{paqbdef}
Q_B&=&2\,{\rm Tr}_A[\rho_{AB}\, (P_A\otimes \mathbbm{1}_2)]\nonumber \\ 
&=&\frac{1}{2}\, \sum_{\nu}\, \left( \Lambda^T\,\mathbf{p_A} \right)_{\nu}\,  \, \sigma_{\nu}
\end{eqnarray} 
which results in the Minkowski four-vector transformation 
\begin{equation} 
\label{pos1}
\mathbf{q}_B=\Lambda^T\,\mathbf{p}_A.
\end{equation} 
Thus, the map $P_A \mapsto Q_B$ is found to be identical to the four-vector map  $\Lambda^T: \mathbf{p}_A\mapsto \mathbf{q}_B=\Lambda^T\,\mathbf{p}_A$. Non-negativity of  squared Minkowski norm of the four-vector  $\mathbf{q}_B$ ( which corresponds to   $Q_B\geq 0$) leads to      
\begin{eqnarray}
\label{pos2}
\mathbf{q}^T_B\, G\, \mathbf{q}_B\geq 0 &&\Longrightarrow  \ 
\mathbf{p}^T_A\,\Lambda\, G\, \Lambda^T\, \mathbf{p}_A\geq 0  \nonumber \\ 
&&\Longrightarrow \  \mathbf{p}^T_A\,\Omega_{A}\, \mathbf{p}_A\geq 0
\end{eqnarray}  	
where 
\be
\label{oA}
\Omega_{A}=\Lambda\, G\, \Lambda^T
\ee
denotes a real symmetric $4\times 4$ matrix, associated with the real parametrization $\Lambda$ of the two-qubit density matrix $\rho_{AB}.$  
Furthermore, positivity of the zeroth component of the four-vector $\mathbf{p}_A$ imposes that  
\begin{equation}
\label{pa0pos}
p_{A_0}>0 \, \Longrightarrow q_{B_0}=\left(\Lambda^T\,\mathbf{p}_A\right)_0> 0.
\end{equation}

Similarly, the map 
\begin{eqnarray}
\label{mapB2A}
P_B \mapsto Q_A&=& 2\, {\rm Tr}_B\left[(\mathbbm{1}_2\otimes \sqrt{P_B})\, \rho_{AB}\,(\mathbbm{1}_2\otimes \sqrt{P_B})\right] \nonumber \\
&=& 2\,{\rm Tr}_B[\rho_{AB}\, (\mathbbm{1}_2\otimes P_B)] 
\end{eqnarray}
from  the set of all non-negative operators ${\mathcal{P}}_B^+~:=~\{P_B\, \vert\ P_B\geq 0 \}$ acting on the Hilbert space ${\mathcal{H}}_B$  to the set ${\mathcal{Q}}^+_A:=\, \{Q_A\, =2\,  {\rm Tr}_B[\rho_{AB}\, (\mathbbm{1}_2\otimes P_B)]\subset\mathcal{H}_A$ on  the Hilbert space ${\mathcal{H}}_A$ leads to the identification
\begin{eqnarray}
\label{pbqadef}
	 Q_A &=&\frac{1}{2}\,\sum_{\mu}\, \left(\Lambda\,\mathbf{p_B}\right)_{\mu}\,\sigma_{\mu}.
	\end{eqnarray} 
In turn, we obtain the Minkowski four-vector transformation
\begin{equation}
\label{qpos2} 
\mathbf{q}_A=\Lambda\,\mathbf{p}_B.
\end{equation} 
where the four-vector $\mathbf{q}_A$ characterizes a non-negative operator  $Q_A\in {\mathcal{Q}}^+_A$ faithfully. The map $P_B \mapsto Q_A$  reflects itself in terms of the four-vector transformation  $\Lambda~:~\mathbf{p}_B~\mapsto~\mathbf{q}_A=\Lambda\,\mathbf{p}_B$ in the Minkowski space such that 
 \begin{eqnarray}
\label{pos3}
q_{A_0}> 0\Longrightarrow&& \left(\mathbf{\Lambda\, p}_B\right)_0> 0,  \\
 \label{pos4}
\mathbf{q}^T_A\, G\, \mathbf{q}_A\geq 0 \Longrightarrow && \ \mathbf{p}^T_B\,\Lambda^T\, G\, \Lambda\, \mathbf{p}_B\geq 0   \nonumber \\  
\Longrightarrow&&  \ \mathbf{p}^T_B\,\Omega_{B}\, \mathbf{p}_B\geq 0, 
\end{eqnarray}  	 
where, we have denoted
\begin{equation}
\label{oB} 
\Omega_{B} = \Lambda^T\, G\, \Lambda.
\end{equation}

The $4\times 4$ real symmetric matrices $\Omega_{A}=\Lambda\, G\, \Lambda^T$, $\Omega_{B}~=~\Lambda^T\, G\, \Lambda$ constructed from the real counterpart $\Lambda$ of the two-qubit density matrix $\rho_{AB}$ play a central role in our analysis.    
  
\section{Lorentz singular value decomposition of $\Lambda$ and canonical forms of two-qubit density matrix under SLOCC}

From the properties of the $4\times 4$ real matrix $\Lambda$, parametrizing the two-qubit density matrix $\rho_{AB}$, it is clear that (i) under the map 
$\mathbf{p}_{A}\mapsto \mathbf{q}_B=\Lambda^T\,\mathbf{p}_A$  
and $\mathbf{p}_{B}\mapsto \mathbf{q}_A=\Lambda\,\mathbf{p}_B$ ( (see (\ref{pos1}), 
(\ref{qpos2})),  four-vectors $\mathbf{p}_{A}$, $\mathbf{q}_{A}$   with Minkowski norms $\mathbf{p}_A^T\,G\mathbf{p}_A\geq 0$, $\mathbf{q}_A^T\,G\mathbf{q}_A\geq 0$  and  positive zeroth components $p_{A_0}> 0$, $q_{A_0}> 0$  get    transformed to  four-vectors $\mathbf{q}_B$, $\mathbf{p}_B$   such that  $\{\mathbf{q}_B^T\,G\mathbf{q}_B\geq 0,\ q_{B_0}> 0\}$,  $\{\mathbf{p}_B^T\,G\mathbf{p}_B\geq 0,\ p_{B_0}> 0\}$  respectively. Furthermore   (ii) the sets  $\left\{\Lambda\, \mathbf{p}\right\vert  \mathbf{p}^T\, G\,\mathbf{p}\geq 0,   p_0> 0 \}$  and $\left\{\tilde{\mathbf{p}}=L_A\, \Lambda\,L_B^T\,  \mathbf{p}\right\vert  \tilde{\mathbf{p}}^T\, G\,\tilde{\mathbf{p}}\geq 0,   \tilde{p}_0> 0 \}$  are equivalent, as they are related to each other via SLOCC  
transformations on the two-qubit state $\rho_{AB}$. 

Our interest is to look for a particularly simple canonical form, as in   (\ref{can}) for  $\Lambda$, by identifying  suitable OPLG transformations $L_{A_c}, \, L_{B_c}$.  In terms of the real, symmetric matrices $\Omega_A=\Lambda\, G\, \Lambda^T$ and  $\Omega_B=\Lambda^T\, G\, \Lambda$ introduced in  (\ref{oA}),(\ref{oB}), we express,   
\begin{eqnarray}
\label{oAc}
\Omega^c_A&=&L_{A_c}\, \Lambda\, L_{B_c}^T\, G \, L_{B_c}\, \Lambda^T\, L_{A_c}^T \nonumber \\ 
&=& L_{A_c}\, \Omega_A\, L_{A_c}^T    \\ 
\label{oBc}
\Omega^c_B&=& L_{B_c}\, \Lambda^T\, L_{A_c}^T\, G \, L_{A_c}\, \Lambda\, L_{B_c}^T \nonumber \\ 
&=& L_{B_c}\, \Omega_B\, L_{B_c}^T
\end{eqnarray}
where we have used the defining property $L^T\, G\, L=G$ of Lorentz transformation matrix $L$ and denoted the canonical forms of the real, symmetric matrices  $\Omega_A,\  \Omega_B$ by $\Omega^c_A$, $\Omega^c_B$  respectively. 
 
We would like to emphasize here that the canonical form $\Omega^{c}_{A}$ is determined completely  by the real matrix $\Lambda$ and the OPLG tranformations $L_{A_{c}}$ (see (\ref{oAc})). Similarly, $\Omega^{c}_{B}$ is entirely characterized by $\Lambda$ and $L_{B_{c}}$ (see (\ref{oBc})). Therefore, it is possible to introduce the following  canonical forms $\Lambda^c_A$, $\Lambda^c_B$ (upto normalization) for the real matrix $\Lambda$, associated with  $\Omega^c_A$ and $\Omega^c_B$ respectively:      
\begin{eqnarray}
\label{lABc}
\Lambda^c_A= L_{A_c}\, \Lambda\, L_B^T,\ \ \Lambda^c_B= L_{B_c}\, \Lambda^T\, L_A^T. 
\end{eqnarray}
Note that in (\ref{lABc})  the OPLG transformations $L_{A_c}$, $L_{B_c}$   correspond  to physical SLOCC operations carried out by Alice, Bob on their parts of the two-qubit state; but the operations $L_A,\, L_B$ denote any arbitrary OPLG transformations, which respectively leave the structure of $\Omega^c_A$, $\Omega^c_B$ unaltered. Thus,  we express          
\begin{eqnarray}
\label{OABc2}
\Omega^c_A=\Lambda^c_{A}\, G\,  \left(\Lambda^c_{A}\right)^T,\ \ \Omega^c_B= \Lambda^c_{B}\, G\,  \left(\Lambda^c_{B}\right)^T 
\end{eqnarray}
by substituting (\ref{lABc}). 
 
Continuing further, we note  that the Lorentz congruent transformations 
\begin{eqnarray}
\label{oplgab}
\Omega_{A}\longrightarrow\widetilde{\Omega}_{A}=L_{A}\, \Omega_{A}\, L_{A}^T,\nonumber  \\ 
\Omega_{B}\longrightarrow   \widetilde{\Omega}_{B}=L_{B}\, \Omega_{B}\, L_{B}^T  
\end{eqnarray}
are not   similarity transformations. But, the following  pair of matrices   
\begin{eqnarray}
\label{goab}
G\, \Omega_{A}= G\,\Lambda\, G\, \Lambda^T, \ \ 
G\, \Omega_{B}= G\,\Lambda^T\, G\, \Lambda  
\end{eqnarray}
do undergo similarity transformations    
\begin{eqnarray}
\label{goas}
G\, \Omega_{A} \, &\longrightarrow&\, G\,  L_{A}\,  \Omega_{A} \, L_{A}^T  \nonumber  \\
&=& \left(L^T_{A}\,\right)^{-1}\, G\,\Omega_{A}\, L_{A}^T  \\
\label{gobs}
G\, \Omega_{B}\, &\longrightarrow&\, G\,  L_{B}\,  \Omega_{B} \, L_{B}^T  \nonumber  \\
&=& \left(L^T_{B}\,\right)^{-1}\, G\,\Omega_{B}\, L_{B}^T, 
\end{eqnarray}
when $\Omega_{A},\, \Omega_{B}$ undergo OPLG transformations  (\ref{oplgab})). In  (\ref{goas}), (\ref{gobs}),  we have used $G\, L\,=\left(L^T\right)^{-1}\, G$ satisfied by every OPLG matrix $L$.  It is evident that the eigenvalues of 
$G\, \Omega_{A}$ and $G\, \Omega_{B}$ remain invariant under OPLG transformations $L_A$, $ L_B$, associated with the SLOCC operations on qubits $A$ and $B$ respectively. Furthermore, it is readily seen that the eigenvalues of  $G\, \Omega_{A}$, $G\, \Omega_{B}$  are identical as
\begin{equation}
{\rm Tr}\,[(G\,\Omega_{A})^n]={\rm Tr}\,[(G\,\Omega_{B})^n],\ n=1,2,\ldots.  
\end{equation}
Based on a detailed algebraic analysis  carried out by  some of us~\cite{AVG1,AVG2,note2} on $4\times 4$ real matrices, satisfying the conditions (\ref{pos2}), (\ref{pa0pos}), (\ref{pos3}), (\ref{pos4}), we state the following theorem on the nature of eigenvalues and eigenvectors of the matrices $G\,\Omega_{A}$  ($G\,\Omega_{B}$):  

\noindent {\bf Theorem}: The  $4\times 4$ real matrix  $G\,\Omega_{A}$ ($G\, \Omega_{B}$) associated with the  real form  $\Lambda$ of a two-qubit density matrix $\rho_{AB}$ necessarily possesses 
\begin{itemize}
 \item[(i)] non-negative eigenvalues; 
 \item [(ii)]either {\em positive} or {\em neutral} eigenvector corresponding to its highest eigenvalue;
\item [(iii)] a set of eigenvectors consisting of either 
\begin{itemize}
 \item[(a)] {\bf one}  positive four-vector belonging to the highest eigenvalue and {\bf three}  negative four-vectors  \\
 or 
 \item [(b)] {\bf one}  neutral four-vector belonging to  the highest eigenvalue --  atleast doubly degenerate -- and  {\bf two}  negative four-vectors.  
\end{itemize}
 \end{itemize}

From the above theorem (see Appendix A for a concise proof)  it follows that two different cases arise, depending on whether the eigenvector belonging to the highest eigenvalue of $G\,\Omega_{A}$ ($G\,\Omega_{B}$) is {\em positive} or {\em neutral}. Consequently, we have two types of canonical forms for 
$\Omega_{A}$ ($\Omega_{B}$) and consequently, for the real parametrization  $\Lambda$,  the corresponding two-qubit density matrix $\rho_{AB}$ under SLOCC transformations. Note that the eigenvalues, eigenvectors of  $G\,\Omega_{A}$ ($G\,\Omega_{B}$) are also referred to as G-eigenvalues and G-eigenvectors of  the real symmetric matrix $\Omega_{A}$ ($\Omega_{B}$).  

Next, we proceed to find two different types of canonical forms of the real symmetric matrices $\Omega_{A}$, $\Omega_{B}$.


\subsection{Type-I canonical form} 

Let us arrange the eigenvalues  of  $G\,\Omega_A$, $G\, \Omega_B$   in the  order $\lambda_0\geq \lambda_1\geq \lambda_2\geq\lambda_3$ and denote the associated set of eigenvectors  by   $\{\mathbf{a}_0,\, \mathbf{a}_1,\, \mathbf{a}_2,\, \mathbf{a}_3\}$ and $\{\mathbf{b}_0,\, \mathbf{b}_1,\, \mathbf{b}_2,\, \mathbf{b}_3\}$, respectively.  Suppose that $\mathbf{a}_0$ and $\mathbf{b}_0$ are positive four-vectos. 
From (iii\, a) of the theorem, it is clear that the set of eigenvectors $\{\mathbf{a}_0,\, \mathbf{a}_1,\, \mathbf{a}_2,\, \mathbf{a}_3\}$ and $\{\mathbf{b}_0,\, \mathbf{b}_1,\, \mathbf{b}_2,\, \mathbf{b}_3\}$ associated with the eigenvalues $\lambda_0, \lambda_1,\lambda_2,\lambda_3$ of the respective matrices $G\,\Omega_A$, $G\, \Omega_B$ form Minkowski G-orthoronal tetrads (see Appendix B for details) obeying  
\begin{eqnarray}
\label{GorthA}
\mathbf{a}^T_{\mu}\, G\, \mathbf{a}_{\nu}&=&G_{\mu\,\nu} \\ 
\label{GorthB}
\mathbf{b}^T_{\mu}\, G\, \mathbf{b}_{\nu}&=&G_{\mu\,\nu} 
\end{eqnarray}
where \, $\mu,\,\nu=0,1,2,3,$ and $G_{\mu\,\nu}$ are elements of the Minkowski matrix $G$. 
We  construct   OPLG canonical transformation matrices   $L^T_{A_{{\rm I}_c}}$, $L^T_{B_{{\rm I}_c}}$ explicitly (see Appendix B),  
\begin{eqnarray}
\label{la}
L^T_{A_{{\rm I}_c}} 
&=& \left(\begin{array}{llll} \mathbf{a}_{0} &  \mathbf{a}_{1} & \mathbf{a}_{2} & \mathbf{a}_{3} \end{array}\right),  \\ 
\label{lb}
L^T_{B_{{\rm I}_c}} 
&=& \left(\begin{array}{llll} \mathbf{b}_{0} &  \mathbf{b}_{1} & \mathbf{b}_{2} & \mathbf{b}_{3} \end{array}\right)  
\end{eqnarray}
by arranging the  eigenvectors   $\{\mathbf{a}_0,\, \mathbf{a}_1,\, \mathbf{a}_2,\, \mathbf{a}_3\}$ and $\{\mathbf{b}_0,\, \mathbf{b}_1,\, \mathbf{b}_2,\, \mathbf{b}_3\}$  of  $G\,\Omega_A$, $G\,\Omega_B$  as columns of $L^T_{A_{{\rm I}_c}}$, $L^T_{B_{{\rm I}_c}}$ respectively.   

Using (\ref{GorthA}), (\ref{GorthB}), (\ref{la}), (\ref{lb}) and the property 
\begin{eqnarray}
\label{evga}
G\,\Omega_A\, \mathbf{a}_\mu\, &=&\, \lambda_\mu\, \mathbf{a}_\mu 
\, \Rightarrow  \Omega_A\, \mathbf{a}_\mu\, = \, \lambda_\mu\, G\, \mathbf{a}_\mu \\
\label{evgb}
G\,\Omega_B\, \mathbf{b}_\mu\, &=&\, \lambda_\mu\, \mathbf{b}_\mu 
\, \Rightarrow  \Omega_B\, \mathbf{b}_\mu\, = \, \lambda_\mu\, G\, \mathbf{b}_\mu
\end{eqnarray}       
of the eigenvectors of $G\,\Omega_A,\, G\,\Omega_B$,  we arrive at  the diagonal canonical  forms   $\Omega_{A_c}, \Omega_{B_c}$: 
\begin{eqnarray}
\Omega^{{\rm I}_c}_{A}&=& L_{A_{{\rm I}_c}} \Omega_A   L^T_{A_{{\rm I}_c}}=
\left(\begin{array}{cccc}
\label{omegaabc}
\lambda_0 & 0 & 0 & 0 \\ 
0 &  -\lambda_1 & 0 & 0 \\ 
0 & 0 &   -\lambda_2 & 0 \\
0 & 0 & 0 & -\lambda_3
\end{array}\right)    \nonumber  \\
\Omega^{{\rm I}_c}_{B}\,&=&  L_{B_{{\rm I}_c}}  \Omega_B L^T_{B_{{\rm I}_c}} = 
\left(\begin{array}{cccc}
\lambda_0 & 0 & 0 & 0 \\ 
0 &  -\lambda_1 & 0 & 0 \\ 
0 & 0 &   -\lambda_2 & 0 \\
0 & 0 & 0 & -\lambda_3
\end{array}\right).    \nonumber \\ 
\end{eqnarray}

\noindent{\bf Corollary~1:} Under the canonical OPLG transformations $L_{A_{\rm{I}_c}}, \, L_{B_{\rm{I}_c}}$ 
as in (\ref{la}), (\ref{lb}), the real matrix $\Lambda$, with ${\rm sgn}\left( \det(\Lambda)\right)=\pm$, assumes the following diagonal canonical forms:  
\begin{eqnarray} 
\label{diagAIc}
\Lambda^{\rm{I}_c}_A&=& \frac{L_{A_{\rm{I_c}}}\, \Lambda\, L^T_{B}}
{\left(L_{A_{\rm{I_c}}}\, \Lambda\, L^T_{B}\right)_{00}} \nonumber \\  
 &=&{\rm diag}  \,  \left(1,\sqrt{\frac{\lambda_1}{\lambda_0}},\sqrt{\frac{\lambda_2}{\lambda_0}},\, 
\pm\, \sqrt{\frac{\lambda_3}{\lambda_0}}\right), \nonumber \\
&& \\  
\Lambda^{\rm{I}_c}_B&=& \frac{L_{A}\, \Lambda\, L^T_{B_{\rm{I_c}}}}
{\left(L_{A}\, \Lambda\, L^T_{B_{\rm{I_c}}}\right)_{00}}  \nonumber \\
 &=&{\rm diag}  \,  \left(1,\sqrt{\frac{\lambda_1}{\lambda_0}},\sqrt{\frac{\lambda_2}{\lambda_0}},\, 
\pm\, \sqrt{\frac{\lambda_3}{\lambda_0}}\right), \nonumber \\  
&&\ \ \ \ \ \ \lambda_0\geq \lambda_1\geq \lambda_2\geq \lambda_3 \nonumber
\end{eqnarray}
if and only if the eigenvectors corresponding to the highest eigenvalue $\lambda_0$ of   $G\,\Omega_{A}$, $G\, \Omega_{B}$ are positive four-vectors in ${\mathcal{M}}$.

\noindent {\bf Proof:} It follows from explicit evaluation that 
\begin{eqnarray*}
	\Omega^{{\rm I}_c}_A = \lambda_0\,   \Lambda^{{\rm I}_c}_A G \left(\Lambda^{{\rm I} _c}_A\right)^T, \\ 
\Omega^{{\rm I}_c}_B= \lambda_0\,   \left(\Lambda^{{\rm I}_c}_B\right)^T G \left(\Lambda^{{\rm I}_c}_B\right).  
\end{eqnarray*}
\hskip 2.99in $\square$

Expressed in terms of the three term  factorization $$\Lambda=\left(L_{A_{{\rm I}_c}}\right)^{-1}\,\Lambda^{{\rm I}_c}_A\,\left(L^T_B\right)^{-1}$$  (or equivalently  $\Lambda=\left(L_{B_{{\rm I}_c}}\right)^{-1}\,\Lambda^{{\rm I}_c}_B\,\left(L^T_A\right)^{-1}$), it is evident that  $\Lambda$ is characterized by 15 real parameters, where 6 real parameters  each are from the  Lorentz transformations $L_{A_{{\rm I}_c}}$, $L_{B}$  (or $L_{B_{{\rm I}_c}}$, $L_{A}$)    and the rest of the three  real parameters are given by   $\sqrt{\lambda_i/\lambda_0},\ i=1,2,3$.      

It is easy to see that the two-qubit density matrix  $\rho_{AB}^{\rm{I}_c}$ associated with 
both the canonical forms $\Lambda^{\rm{I}_c}_A,\ \Lambda^{\rm{I}_c}_B$ is a  Bell-diagonal state: 
 \begin{eqnarray}
\label{bds}
 \rho_{AB}^{\rm{I}_c}&=&\frac{1}{4}\left(\mathbbm{1}_2\otimes \mathbbm{1}_2+\,  \sum_{i=1,2}\, \sqrt{\frac{\lambda_i}{\lambda_0}}\, \sigma_i\otimes \sigma_i \right. \nonumber \\
 &&\hskip 0.75in \left. \mp\sqrt{\frac{\lambda_3}{\lambda_0}}\, \sigma_3\otimes \sigma_3\right) \\ 
 &=&  \rho_{BA}^{\rm{I}_c}. \nonumber
\end{eqnarray}

\subsection{Type-II canonical forms}
Suppose the maximum eigenvalue $\lambda_0$ of $G\,\Omega_{A}$, associated with the neutral eigenvector 
$\mathbf{u}_0$ is atleast doubly degenerate. Let us denote the set of eigenvalues of $G\,\Omega_{A}$  by 
$\{\lambda_0,\ \lambda_0,\ \lambda_1,\ \lambda_2\,\}$ arranged as $\lambda_0\geq \lambda_1\geq\lambda_2$. From (iii b) of the theorem, we have a maximal G-orthogonal triad $\{\mathbf{u}_0,\, \widetilde{\mathbf{a}}_1,\, \widetilde{\mathbf{a}}_2\}$ of eigenvectors  of  $G\,\Omega_{A}$ obeying   
\begin{eqnarray}
\label{GorthA2}
\begin{array}{ll} \mathbf{u}_0^T\, G\, \mathbf{u}_0=0, &  \mathbf{u}_0^T\, G\, \widetilde{\mathbf{a}}_i=0, \\
\widetilde{\mathbf{a}}_i^T\, G\, \widetilde{\mathbf{a}}_j=-\delta_{i\,j}, &  \ i,\,j=1,\,2 \end{array}   \nonumber 
	\end{eqnarray}
As outlined in the Appendix B we construct a  G-orthogonal tetrad $\{\widetilde{\mathbf{a}}_0,\, \widetilde{\mathbf{a}}_1,\, \widetilde{\mathbf{a}}_2,\, \widetilde{\mathbf{a}}_3\}$ of four-vectors from the given set $\{\mathbf{u}_0,\, \widetilde{\mathbf{a}}_1,\, \widetilde{\mathbf{a}}_2\}$ of the eigenvectors of $G\Omega_A$ which consists of one neutral and two negative four-vectors. 

Chosing a four-vector $\mathbf{u}_3$ such that $\mathbf{u}_3^T\, G\, \mathbf{u}_0\neq 0$ and  $\mathbf{u}_3^T\, G\,\widetilde{\mathbf{a}}_i=0,\ i=1,2$, we construct  
\begin{eqnarray}   
\label{atilde}
\widetilde{\mathbf{a}}_0&=& \mathbf{u}_3 + \tau_u\, \mathbf{u}_0,\ \  \widetilde{a}_{00}\geq 0  \nonumber \\
& & \\
\widetilde{\mathbf{a}}_3 &=& \mathbf{u}_3 - \kappa_u\, \mathbf{u}_0  \nonumber 
\end{eqnarray}
where  
\begin{eqnarray} 
\label{tauu}
\tau_{u}&=& \frac{1-\mathbf{u}_3^T\, G\, \mathbf{u}_3}{2\, \mathbf{u}_3^T\, G\, \mathbf{u}_0},\ \ \nonumber  \\ 
& & \\
\kappa_u&=& \frac{1+\mathbf{u}_3^T\, G\, \mathbf{u}_3}{2\, \mathbf{u}_3^T\, G\, \mathbf{u}_0}. \nonumber
\end{eqnarray}
The tetrad $\{\widetilde{\mathbf{a}}_0,\widetilde{\mathbf{a}}_1,\widetilde{\mathbf{a}}_2,\widetilde{\mathbf{a}}_3\}$ of four-vectors satisfy the G-orthogonality conditions: $\widetilde{\mathbf{a}}^T_\mu\, G\, \widetilde{\mathbf{a}}_\nu=G_{\mu\nu}$. 

On arranging the G-orthogonal tetrad $\{\widetilde{\mathbf{a}}_0, \widetilde{\mathbf{a}}_1,\widetilde{\mathbf{a}}_2,\widetilde{\mathbf{a}}_3\}$ as columns, we construct the OPLG matrix    
\begin{eqnarray}
\label{la2}
L_{A_{{\rm{II}_c}}}&=&\left(\begin{array}{llll} \widetilde{\mathbf{a}}_{0} &  \widetilde{\mathbf{a}}_{1} & \widetilde{\mathbf{a}}_{2} & \widetilde{\mathbf{a}}_{3} \end{array}\right), 
\end{eqnarray}
in order to transform  $\Omega_{A}$ to its canonical form.   

Let us denote the `00' element of the matrix $\Omega^{{\rm II}_c}_{A}$  by 
\begin{eqnarray}
\label{p0}
\phi_0&=&\left(L_{A_{{\rm II}_c}}\,\Omega_A\, L^{T}_{A_{{\rm II}_c}}\right)_{00} \nonumber \\
&=& \widetilde{\mathbf{a}}_0^T\, \Omega_{A}\, \widetilde{\mathbf{a}}_0.
\end{eqnarray}
Substituting  (\ref{atilde}),  (\ref{tauu}), (\ref{p0})  and simplifying `30' and `33'  matrix elements of $\Omega^{{\rm II}_c}_{A}$ we obtain 
\begin{eqnarray}
\label{pgeq0}
\left(L_{A_{{\rm II}_c}}\,\Omega_A\, L^{T}_{A_{{\rm II}_c}}\right)_{30}&=&\widetilde{\mathbf{a}}_3^T\, \Omega_A\, \widetilde{\mathbf{a}}_0 \nonumber \\
&=& \phi_0-\lambda_0 \nonumber \\ 
&& \\
\left(L_{A_{{\rm II}_c}}\,\Omega_A\, L^{T}_{A_{{\rm II}_c}}\right)_{33}&=&\widetilde{\mathbf{a}}_3^T\, \Omega_A\, \widetilde{\mathbf{a}}_3 \nonumber \\
&=& \phi_0-2\, \lambda_0. \nonumber
\end{eqnarray}
We thus arrive at the non-diagonal type-II canonical form of the real symmetric matrix $\Omega_{A}$ as 
\begin{eqnarray}
\Omega^{\rm{II}_c}_{A}&=& L_{A_{{\rm{II}_c}}} \Omega_{A}  L^T_{A_{{\rm{II}_c}}}\nonumber \\ 
&=&
\left(\begin{array}{cccc}
\phi_0 & 0  & 0 & \phi_0-\lambda_0 \\ 
0 & -\lambda_1 & 0 & 0 \\ 
0 & 0 &   -\lambda_2 & 0 \\
\phi_0-\lambda_0 & 0 & 0 &  \phi_0-2\,\lambda_0
\end{array}\right)  
\end{eqnarray}
where   $\lambda_0 \geq \lambda_1\geq \lambda_2$. 

In an analogous manner, we consider the  triad   $\{\mathbf{v}_0,\, \widetilde{\mathbf{b}}_1,\, \widetilde{\mathbf{b}}_2\}$ of eigenvectors of $G\,\Omega_{B}$ corresponding respectively to the eigenvalues $\lambda_0$ (doubly degenerate), $\lambda_1$ and $\lambda_2$. The eigenvectors satisfy G-orthogonality conditions       
\begin{eqnarray}
\label{GorthB2}
\begin{array}{ll} \mathbf{v}_0^T\, G\, \mathbf{v}_0=0, & \mathbf{v}_0^T\, G\, \widetilde{\mathbf{b}}_k=0, \\
\widetilde{\mathbf{b}}_k^T\, G\, \widetilde{\mathbf{b}}_l=-\delta_{k\,l}, & \ k,l=2,3. \end{array}   
\end{eqnarray}
Starting from this eigenvector set containing one neutral and two negative four-vectors, we pick a four-vector $\mathbf{v}_3$ such that $\mathbf{v}_3^TG\, \mathbf{v}_0\neq 0$ and  $\mathbf{v}_3^TG\,\widetilde{b}_i=0,\ i=1,2$, and construct   (see Appendix B for details)    
\begin{eqnarray}   
\label{btilde}
\widetilde{\mathbf{b}}_0&=& \mathbf{v}_3 + \tau_v\, \mathbf{v}_0,\ \ \ \widetilde{b}_{00} \geq 0  \nonumber \\
&& \\
\widetilde{\mathbf{b}}_3 &=& \mathbf{v}_3 - \kappa_v\, \mathbf{v}_0 \nonumber 
\end{eqnarray}
where 
\begin{eqnarray} 
\label{taukappav} 
\tau_{v}&=& \frac{1-\mathbf{v}_3^T\, G\, \mathbf{v}_3}{2\, \mathbf{v}_3^T\, G\, \mathbf{v}_0},\ \nonumber  \\
& & \\
\kappa_v&=& \frac{1+\mathbf{v}_3^T\, G\, \mathbf{v}_3}{2\, \mathbf{v}_3^T\, G\, \mathbf{v}_0}. \nonumber 
\end{eqnarray} 
This helps us to identify the tetrad $\{\widetilde{\mathbf{b}}_0, \widetilde{\mathbf{b}}_1, \widetilde{\mathbf{b}}_2, \widetilde{\mathbf{b}}_3 \}$ of four-vectors obeying G-orthogonality conditions
$\widetilde{\mathbf{b}}^T_\mu\, G\, \widetilde{\mathbf{b}}_\nu=G_{\mu\nu}$. So, we can explicitly  construct the canonical OPLG matrices  
\begin{eqnarray}
\label{lb2}
L^T_{B_{{\rm II}_c}} &=&  
 \left(\begin{array}{llll} \widetilde{\mathbf{b}}_{0} &  \widetilde{\mathbf{b}}_{1} & \widetilde{\mathbf{b}}_{2} & \widetilde{\mathbf{b}}_{3} \end{array}\right).  
\end{eqnarray} 
and obtain the canonical form of the real symmetric matrix $\Omega_{B}=\Lambda^T\, G\, \Lambda$ as,  
\begin{eqnarray}
\label{OmegabIIc}
\Omega^{\rm{II}_c}_{B}&=& L_{B_{{\rm{II}_c}}} \Omega_{B}  L^T_{B_{{\rm{II}_c}}}\nonumber \\ 
&=&
\left(\begin{array}{cccc}
\chi_0 & 0  & 0 & \chi_0-\lambda_0 \\ 
0 & -\lambda_1  & 0 & 0 \\ 
0 & 0 &   -\lambda_2 & 0 \\
\chi_0-\lambda_0 & 0 & 0 & \chi_0-2\,\lambda_0
\end{array}\right).  
\end{eqnarray} 
Here, we have denoted the `00' element of $\Omega^{{\rm II}_c}_B$ by    by   
\begin{equation}
\label{q0}
\chi_0=\left(L_{B_{{\rm II}_c}}\,\Omega_B\, L^{T}_{B_{{\rm II}_c}}\right)_{00}=\widetilde{\mathbf{b}}_0^T\, \Omega_{B}\, \widetilde{\mathbf{b}}_0.
\end{equation}  
Then, we evaluate  `30' and '33' elements of of $\Omega^{{\rm II}_c}_B$   
by substituting  (\ref{btilde}), (\ref{taukappav}), (\ref{q0}) to obtain  
\begin{eqnarray}
\label{qgeq0}
\left(L_{B_{{\rm II}_c}}\,\Omega_B\, L^{T}_{B_{{\rm II}_c}}\right)_{30}
&=& \widetilde{\mathbf{b}}_3^T\, \Omega_B\, \widetilde{\mathbf{b}}_0 \nonumber \\ 
&=& \chi_0-\lambda_0 \nonumber \\ 
&& \\
\left(L_{B_{{\rm II}_c}}\,\Omega_B\, L^{T}_{B_{{\rm II}_c}}\right)_{33}
&=& \widetilde{\mathbf{b}}_3^T\, \Omega_B\, \widetilde{\mathbf{b}}_3 \nonumber \\ 
&=&\chi_0-2\,\lambda_0. \nonumber  
\end{eqnarray}

\noindent{\bf Corollary~2:} When the eigenvectors corresponding to -- at least doubly degenerate -- largest eigenvalue $\lambda_0$ of the matrix $G\,\Omega_{A}$ ($G\,\Omega_{B}$) is a neutral four-vector in ${\mathcal{M}}$, there exist  canonical OPLG transformations $L_{A_{{\rm{II}_c}}}, \, L_{B_{{\rm{II}_c}}}$ yielding the following  {\emph{non-diagonal}} canonical forms of the real matrix $\Lambda$,  
\begin{eqnarray} 
\label{aIIcan}
\Lambda_{A}^{\rm {II}_c} &=&  \frac{
L_{A_{{\rm II}_c}} \Lambda\, L_B^T}
{\left(L_{A_{{\rm II}_c}} \Lambda\, L_B^T\right)_{00}} \nonumber \\
&=& \left(\begin{array}{cccc}
1 & 0 & 0 & 0 \\ 
0 & \sqrt{\frac{\lambda_1}{\phi_0}}  &0 & 0 \\ 
0 & 0 & \pm\sqrt{\frac{\lambda_2}{\phi_0}}  & 0 \\ 
1-\frac{\lambda_0}{\phi_0} & 0 & 0 & \frac{\lambda_0}{\phi_0} 
\end{array}\right)  
\end{eqnarray}
where $L_B$ is an OPLG transformation 
and 
\begin{eqnarray} 
\label{bIIcan}
\Lambda_{B}^{\rm {II}_c} &=&  \frac{L_A\, \Lambda\, L^T_{B_{{\rm II}_c}} 	}
{\left(		L_A\, \Lambda\, L^T_{B_{{\rm II}_c}} \right)_{00}} \nonumber \\
&=& \left(\begin{array}{cccc}
1 & 0 & 0 & 1-\frac{\lambda_0}{\chi_0} \\ 
0 & \sqrt{\frac{\lambda_1}{\chi_0}}  &0 & 0  \\ 
0 & 0 & \pm\sqrt{\frac{\lambda_2}{\chi_0}}  & 0 \\ 
0 & 0 & 0 & \frac{\lambda_0}{\chi_0} 
\end{array}\right)  
\end{eqnarray}
with $L_A$ being a suitable OPLG transformation. 
 
Depending on if ${\rm sgn}\left(\det\Lambda\right)=\pm$, one obtains  `$\pm$' sign in  the diagonal element $\left(\Lambda_{A,\, {\rm or}\,  B }^{\rm {II}_c}\right)_{22}$  in (\ref{aIIcan}), (\ref{bIIcan}).

\noindent{\bf Proof:} It readily follows from explicit evaluations that    
\begin{eqnarray*}
\Omega^{\rm{II}_c}_{A}&=&\phi_0\,  \Lambda_{A}^{\rm{II}_c}\, G\, \left(\Lambda_{A}^{\rm{II}_c}\right)^T
\end{eqnarray*}
and     
\begin{eqnarray*}
	\Omega^{\rm{II}_c}_{B}&=& \chi_0\, \left(\Lambda_{B}^{\rm{II}_c}\right)^T \, G\, \Lambda_{B}^{\rm{II}_c}. 
\end{eqnarray*}
\hskip 3in$\square$ 
		
\noindent {\bf Remark:} From our discussions in  Sec.~2, which resulted in the identification of real symmetric matrices $\Omega_{A}$, $\Omega_B$ associated with  the real parametrization $\Lambda$ of the two-qubit density matrix $\rho_{AB}$ (see (\ref{rho}),(\ref{block}), (\ref{paqbdef})--(\ref{oA}),  (\ref{pbqadef})--(\ref{oB})), we observe that  
\begin{enumerate}[label=(\roman*)]
\item  while the real matrix $\Lambda$ parametrizes the two-qubit density matrix  $\rho_{AB}$,  its transpose $\Lambda^T$  characterizes  $\rho_{BA}$, which is obtained by swapping $A$ and $B$;
\item  canonical SLOCC  transformations  $\rho_{AB}\longrightarrow \rho^c_{AB}$, and 
$\rho_{BA}\longrightarrow \rho^c_{BA}$ are governed by  the eigenvalues and the eigenvectors of the real matrices 
$G\,\Lambda \, G\,  \Lambda^T = G\,\Omega_{A}$, $G\,\Lambda^T\,G\, \Lambda= G\,\Omega_{B}$ respectively; 
\item  even though $G\,\Omega_{A}$, $G\,\Omega_{B}$ share same eigenspectrum, the associated set of eigenvectors is different, in general and hence one may expect  different canonical structures 
$\rho^c_{AB}$, $\rho_{BA}^c$ for the density matrices $\rho_{AB}$, $\rho_{BA}$; 
\item exactly identical canonical forms $\Lambda_A^{{\rm I}_c}=\Lambda_B^{{\rm I}_c}$ (see \ref{diagAIc}) and correspondingly $\rho^c_{AB}=\rho_{BA}^c$ (see (\ref{bds})) are obtained when the eigenvectors of  $G\,\Omega_{A}$, $G\,\Omega_{B}$ corresponding to their highest eigenvalue  are positive four-vectors in $\mathcal{M}$; 
\item when neutral four-vectors in $\mathcal{M}$ happen to be one of the eigenvectors of  $G\,\Omega_{A}$, $G\,\Omega_{B}$ (corresponding to at least doubly repeated highest eigenvalue $\lambda_0$) there are two different OPLG canonical forms  (see  (\ref{aIIcan}), (\ref{bIIcan}))  $\Lambda^{\rm{II}_c}_{A}$, $\Lambda^{\rm{II}_c}_{B}$ ,  and hence, SLOCC canonical forms $\rho_{AB}^{\rm{II}_c}$, $\rho_{BA}^{\rm{II}_c}$ of the corresponding density matrix $\rho_{AB}$ differ, in general;  
\item when  $\Omega^{\rm{II}_c}_{A}=\Omega^{\rm{II}_c}_{B}$ one obtains  $\Lambda_{A}^{\rm{II}_c}=\left(\Lambda_{B}^{\rm{II}_c}\right)^T$.  
\end{enumerate}    

Corresponding to the type-II canonical form $\Lambda_A^{{\rm II}_c}$ given by (\ref{aIIcan}) we obtain explicit matrix form of $\rho_{AB}^{\rm{II}_c}$  (in the standard two-qubit basis $\{\vert 0_A,0_B\rangle, \vert 0_A,1_B\rangle, \vert 1_A,0_B\rangle, \vert 1_A,1_B\rangle\}$):  
\begin{eqnarray} 
\rho_{AB}^{\rm{II}_c}&=&\frac{1}{2}\left(\begin{array}{cccc} 1  & 0 & 0 & \frac{r_1- r_2}{2}  \\  
0 &  (1-r_0)    &   \frac{r_1+r_2}{2} & 0   \\ 
0 & \frac{r_1+r_2}{2} & 0 & 0 \\ 
\frac{r_1-r_2}{2} & 0 & 0 & r_0
\end{array}\right) \nonumber \\
\end{eqnarray}
where we have denoted   
\begin{eqnarray}
	\label{r0}
\frac{\lambda_0}{\phi_0}&=&r_0, \ \ \sqrt{\frac{\lambda_i}{\phi_0}}=r_i,\ \ i=1,2 
\end{eqnarray}
Non-negativity condition $\rho_{AB}^{\rm{II}_c}\geq 0$ of the density matrix  demands that 
\begin{eqnarray}
\label{plambda_a}
r_1= - r_2,\  \ r_0\geq r^2_1. 
\end{eqnarray}

Similarly, explicit matrix structure of the two-qubit density matrix  $\rho^{{\rm II}_c}_{BA}$ associated with  the type-II canonical form $\Lambda^{{\rm II}_c}_B$  (see (\ref{bIIcan}) is given by   
\begin{eqnarray} 
\rho_{BA}^{\rm{II}_c}&=& \frac{1}{2}\left(\begin{array}{cccc} 1  & 0 & 0 & \frac{s_1- s_2}{2}  \\  
0 &  0    &   \frac{s_1+s_2}{2} & 0   \\ 
0 & \frac{s_1+s_2}{2} & (1-s_0) & 0 \\ 
\frac{s_1-s_2}{2} & 0 & 0 & s_0
\end{array}\right) \nonumber \\
\end{eqnarray}
where we have denoted 
\begin{eqnarray}
\label{s0}
	\frac{\lambda_0}{\chi_0}&=&s_0,\ \sqrt{\frac{\lambda_i}{\chi_0}}=s_i, i=1,2. 
\end{eqnarray}		
It is readily seen that  $\rho_{BA}^{\rm{II}_c}\geq 0$ if and only if   
\begin{eqnarray}
\label{plambda_b}
s_1= - s_2,\  s_0\geq s^2_1 
\end{eqnarray}
\begin{widetext}
Substituting  (\ref{plambda_a}), (\ref{plambda_b}), we get {\em bonafide} type-II  Lorentz canonical forms  $\Lambda_{A}^{\rm{II}_c},\  \Lambda_{B}^{\rm{II}_c}$   
and the associated density matrices $\rho_{AB}^{\rm{II}_c},\  \rho_{BA}^{\rm{II}_c}$  as        
\begin{eqnarray} 
\label{abfincanII}
\Lambda_{A}^{\rm{II}_c}=\frac{L_{A_{{\rm II}_c}} \Lambda\, L_B^T}{\left(L_{A_{{\rm II}_c}} \Lambda\, L_B^T\right)_{00}}
&=& \left(\begin{array}{cccc}
1 & 0 & 0 & 0 \\ 
0 & r_1 &0 & 0 \\ 
0 & 0 & -r_1 & 0 \\ 
1-r_0 & 0 & 0 & r_0  
\end{array}\right), \ \ 
\rho_{AB}^{\rm{II}_c}=\frac{1}{2}\left(\begin{array}{cccc} 1  & 0 & 0 &  r_1  \\  
0 &  1-r_0    &  0 & 0   \\ 
0 & 0 & 0 & 0 \\ 
 r_1 & 0 & 0 & r_0
\end{array}\right),\ \ 0\leq r^2_1 \leq r_0\leq 1   \nonumber \\
& & \\ 
\Lambda_{B}^{\rm{II}_c}=\frac{L_A\, \Lambda\,L^T_{B_{{\rm II}_c}}}{\left(L_A\, \Lambda\,L^T_{B_{{\rm II}_c}}\right)_{00}}&=& \left(\begin{array}{cccc}
1 & 0 & 0 & 1-s_0 \\ 
0 & s_1 & 0 & 0 \\ 
0 & 0 & -s_1 & 0 \\ 
0 & 0 & 0 & s_0 
\end{array}\right),  \ \  
\rho_{BA}^{\rm{II}_c}=\frac{1}{2}\left(\begin{array}{cccc} 1  & 0 & 0 &  s_1  \\  
0 &  0    &  0 & 0   \\ 
0 & 0 & 1-s_0 & 0 \\ 
s_1 & 0 & 0 & s_0
\end{array}\right),  \ \ 0\leq s^2_1 \leq s_0\leq 1 \nonumber 
\end{eqnarray}
\end{widetext}     

It is pertinent to point out that  type-II canonical forms are associated with SLOCC transformations on the two-qubit density matrices of rank less than or equal to 3. 
Based on the three term factorization (upto normalization) $\Lambda=\left(L_{A_{{\rm II}_c}}\right)^{-1}\, \Lambda_A^{{\rm II}_c}\, \left(L_B^T\right)^{-1}$,  
 it is clear that the 14 real parameters characterizing  $\Lambda$ are expressed in terms of  12 parameters of the OPLG transformations  $L_{A_{{\rm II}_c}}, L_B$ and the two parameters of the canonical form i.e.,    $r_0=\lambda_0/\phi_0$, $r_1=\sqrt{\lambda_1/\phi_0}$.  (Similarly, $\left(L_{A}\right)^{-1}\, \Lambda_B^{{\rm II}_c}\,\left(L^T_{B_{{\rm II}_c}}\right)^{-1}$) is characterized by  12 real parameters of  transformations $L_{A}, L_{B_{{\rm II}_c}}$ and the canonical parameters   $s_0=\lambda_0/\chi_0$, $s_1=\sqrt{\lambda_1/\chi_0}$).  

\subsection{Non-diagonal SLOCC normal form of  Verstraete et. al.}
Verstraete et. al.~\cite{verstraete2001} had obtained two different types of Lorentz canonical forms of the real matrix $\Lambda$ under the transformation 
$\Lambda\longrightarrow L_A\, \Lambda\, {L}^T_B,\  L_A,\ L_B\in SO(3,1)$,  by making use of theorem (5.3) of Ref.~\cite{Gohberg83} on matrix decompositions in $n$ dimensional  space with indefinite metric. One of the canonical forms of real matrix $\Lambda$ of Ref.~\cite{verstraete2001} is diagonal (type-I) and the corresponding  SLOCC structure of the two-qubit density matrix is Bell-diagonal. Our type-I canonical form (\ref{diagAIc}) for the real matrix $\Lambda$  agrees identically with this result given by  Ref.~\cite{verstraete2001}.  The non-diagonal canonical form of the real matrix $\Lambda$, corresponding to two-qubit states of rank less than four,  has the following explicit structure~\cite{verstraete2001}:  
\begin{eqnarray}
\label{vc}
\Sigma&=& \left(\begin{array}{cccc} 1 & 0 &0 &b \\ 
0 &  d & 0 & 0  \\ 
0 & 0 & -d & 0 \\ 
c & 0 & 0 & 1+c-b  \end{array}\right)		
\end{eqnarray}
where $b,\  c,\  d$ are real parameters. The two-qubit density matrix $\rho^\Sigma_{AB}$ associated with the real matrix  $\Sigma$ is given (in the standard two-qubit basis) by  
\begin{equation}
\label{rhovc}
\rho^{\Sigma}_{AB}= \frac{1}{2}\left(\begin{array}{cccc} 1+c & 0 &0 & d\\ 
0 &  0 & 0  & 0  \\ 
0 & 0 & b-c & 0 \\ 
d & 0 & 0 & 1-b
\end{array}\right). 
\end{equation}
It is clearly seen that the eigenvalues of  $\rho^{\Sigma}_{AB}$ are non-negative if 
\begin{eqnarray} 
\label{eigdcv}
 &&(1+c)\,(1-b)\geq d^2, \ \ 0\leq  (b-c)\leq 2, \nonumber \\   
  && -1\leq b,c,d \leq 1.   
\end{eqnarray} 

In order to establish a connection between the non-diagonal form (\ref{vc}) with the type-II canonical forms (\ref{abfincanII})  we evaluate the symmetric $4\times 4$ matrices 
$\Omega_A=\Sigma\, G\,\Sigma^T$ and $\Omega_B=\Sigma^T\, G\,\Sigma$ associated with the non-diagonal canonical form (\ref{vc}), which are given explicitly by  
\begin{eqnarray*}
	\label{Omgaac}
 \Omega^\Sigma_A&=& \Sigma\, G\,\Sigma^T  \nonumber \\ 
	&=&  \left(\begin{array}{cccc} 1-b^2  & 0 &0 & -(1-b)(b-c) \\ 
		0 &  -d^2 & 0 & 0  \\ 
		0 & 0 & -d^2 & 0 \\ 
		-(1-b)(b-c) & 0 & 0 & (1-b)(b-2\,c-1)  \end{array}\right)\nonumber \\ 		
\label{Omgabc}
\Omega^\Sigma_B&=& \Sigma^T\, G\,\Sigma \\ 
&=& \left(\begin{array}{cccc} 1-c^2  & 0 &0 & (b-c)(1+c) \\ 
	0 &  -d^2 & 0 & 0  \\ 
	0 & 0 & -d^2 & 0 \\ 
(b-c)(1+c) & 0 & 0 & (1+c)(2b-c-1)  \end{array}\right).  \nonumber 		
\end{eqnarray*}
Note that  when $b=c$, the symmetric matrices $\Omega^\Sigma_A$, $\Omega^\Sigma_B$ are diagonal and thus one obtains type-I diagonal canonical form (see subsection~3A) for $\Sigma$. Moreover,  for $b=\pm 1$ or $c=\pm 1$ the density matrix (\ref{rhovc}) reduces to a product form $\rho_{AB}^\Sigma=\rho_A\otimes \rho_B$ where $\rho_A$ or $\rho_B$ are pure states. It is easy to see that the eigenvalues of   $G\Omega_A^\Sigma$ and $G\Omega_B^\Sigma$ are zero in the cases $b=\pm 1$ or $c=\pm 1$. We thus confine our attention to  $b\neq c$, $b,c\neq \pm 1$. 

Eigenvalues of $G\Omega^\Sigma_A,\, G\,\Omega^\Sigma_B$ are readily obtained as    
\begin{eqnarray}
\label{lam0}
\lambda_0&=&\lambda_3=(1+c)(1-b),   \nonumber \\ 
\lambda_1&=&\lambda_2 = d^2.
\end{eqnarray}
From the non-negativity constraint $(1+c)\,(1-b)\geq d^2$ on the density matrix $\rho^\Sigma_{AB}$ (see (\ref{eigdcv})) it follows    
that   $\lambda_0$ happens to be the highest eigenvalue   
and the corresponding eigenvectors of $G\Omega^\Sigma_A,\, G\,\Omega^\Sigma_B$  are neutral four-vectors. This confirms that under SLOCC operations on the two-qubit density matrix $\rho^{\Sigma}_{AB}$ of (\ref{rhovc})  the real matrix $\Sigma$ can be transformed to Lorentz canonical forms of type-II  (see (\ref{abfincanII}) and (\ref{r0}), (\ref{s0})), which we denote by  $\Lambda^{{\rm II}_c}_{A_\Sigma}$ or $\Lambda^{{\rm II}_c}_{B_\Sigma}$. Now we proceed further to obtain explicit matrices corresponding to these type-II canonical forms of $\Sigma$. 

We identify that $\Omega^\Sigma_B$ already exhibits a canonical form as given in (\ref{OmegabIIc}) if we substitute 
\begin{equation}
	\chi_0=\left(\Omega^\Sigma_B\right)_{00}=1-c^2.
	\end{equation}
Thus, we recognize that   $L_{B_{{\rm II}_c}}=\mathbbm{1}_{4}$, i.e., a $4\times 4$ identity matrix.  With the help of  an  OPLG  transformation matrix      
	\begin{eqnarray}
	\label{LA}
	L_{A}&=& \ba{cccc} \frac{1}{\sqrt{1-c^2}} & 0 &0 &  \frac{-c}{\sqrt{1-c^2}}\\ 
	0 & 1  & 0  & 0  \\ 
	0 & 0 & 1 & 0 \\ 
	\frac{-c}{\sqrt{1-c^2}}  & 0 & 0 & \frac{1}{\sqrt{1-c^2}} \ea  
	\end{eqnarray}  
we obtain one of the  type-II canonical sturctures (\ref{abfincanII}) for $\Sigma$: 
\begin{eqnarray}
\label{vcanB}
\Lambda^{{\rm II}_c}_{B_\Sigma}&=&\frac{L_{A}\, \Sigma }{\left(L_A\, \Sigma\,  \right)_{00}}\nonumber \\
& =& \left(\begin{array}{cccc} 1 & 0 &0 & \frac{b-c}{1-c} \\ 
0 &  \frac{d}{\sqrt{1-c^2}} & 0 & 0  \\ 
0 & 0 & \frac{-d}{\sqrt{1-c^2}} & 0 \\ 
0 & 0 & 0 & \frac{1-b}{1-c}  \end{array}\right)
\end{eqnarray}
In other words our type-II canonical form $\Lambda^{{\rm II}_c}_{B_\Sigma}$ is Lorentz equivalent to  the real matrix  $\Sigma$ (see (\ref{vc})) of Ref.~\cite{verstraete2001}. 

Following the method outlined in the subsection~3\,B and in the Appendix B, for the construction of explicit OPLG transformation matrix $L_{A_{{\rm II}_c}}$, we obtain 
	\begin{eqnarray}
\label{LAc}
L_{A_{{\rm II}_c}}&=& \ba{cccc} \frac{1-b+c}{\sqrt{(1+c)(1+c-2b)}} & 0 &0 &  \frac{-b}{\sqrt{(1+c)(1+c-2b)}}\\ 
0 & 1  & 0  & 0  \\ 
0 & 0 & 1 & 0 \\ 
\frac{-b}{\sqrt{(1+c)(1+c-2b)}}  & 0 & 0 & \frac{1-b+c}{\sqrt{(1+c)(1+c-2b)}} \ea  \nonumber \\ 
\end{eqnarray}  
and verify  that 
\begin{eqnarray}
\label{vcanA}
\Lambda^{{\rm II}_c}_{A_\Sigma}&=&\frac{L_{A_{{\rm II}_c}}\, \Sigma }{\left(L_{A_{{\rm II}_c}}\, \Sigma\,  \right)_{00}}\nonumber \\
& =& \left(\begin{array}{cccc} 1 & 0 &0 & 0 \\ 
0 &   \sqrt{\frac{d^2(1+c-2b)}{\lambda_0\,(1-b)}} & 0 & 0  \\ 
0 & 0 &  -\sqrt{\frac{d^2(1+c-2b)}{\lambda_0\,(1-b)}} & 0 \\ 
\frac{c-b}{1-b} & 0 & 0 & \frac{1-2b+c}{1-b}  \end{array}\right)\nonumber \\
\end{eqnarray}
 exhibits type-II canonical form $\Lambda^{{\rm II}_c}_{A}$ given in (\ref{abfincanII}). This proves that the non-diagonal normal form $\Sigma$ (given by (\ref{vc})) is SLOCC equivalent to   the type-II canonical form $\Lambda^{{\rm II}_c}_{A_\Sigma}$ of  (\ref{vcanA})  in confirmity with our approach.     

\section{Geometric representation of SLOCC canonical forms of two-qubits}

It is shown in Sec.~3 that the real matrix $\Lambda$, parametrizing a two-qubit density matrix $\rho_{AB}$, can be reduced to two algebraically distinct types of  canonical forms (\ref{diagAIc}), (\ref{abfincanII}) under OPLG transformations. The algebraically distinct canonical forms are determined via the eigenvalues and eigenvectors of the  matrices 
$G\,\Omega_A~=~G\, \Lambda\,G\,\Lambda^T$ and $G\, \Omega_B~=~G\, \Lambda^T\,G\,\Lambda$ constructed from $\Lambda$ and the Minkowski space metric tensor $G$. In this section we discuss  geometrical representation captured by the canonical forms of  $\Lambda$, which in turn offer  visualization of  the SLOCC invariant families of two-qubit density matrices on and within  the Bloch ball. To this end we recall  (see subsection~2C) that a map  $P_A \mapsto Q_B$ from the set   ${\mathcal{P}}_A^+~:=~\{P_A=\frac{1}{2}\, \sum_{\mu}\, p_{A_\mu}\,\sigma_{\mu}
 \vert\ P_A\geq 0 \}$ of all  non-negative operators acting on the  Hilbert space ${\mathcal{H}}_A$  to another set of non-negative operators ${\mathcal{Q}}^+_B~:=~ \left\{Q_B =2\,  {\rm Tr}_A[\rho_{AB}\, (P_A\otimes \mathbbm{1}_2)]\left\vert  Q_B\geq 0\, \right.\right\}$ on the Hilbert space ${\mathcal{H}}_B$  can be  expressed alternately as a linear transformation  on  Minkowski four-vectors  i.e.,   $\Lambda^T: \mathbf{p}_A\mapsto \mathbf{q}_B=\Lambda^T\,\mathbf{p}_A$, 
 where $\mathbf{p}_A, \mathbf{q}_B$ are non-negative (\,positive/neutral\,) four-vectors with their zeroth components positive $p_{A_0}> 0, \, q_{B_0}> 0$. Similarly, the real matrix $\Lambda$ induces a linear trasnformation $\Lambda: \mathbf{p}_B\mapsto \mathbf{q}_A=\Lambda\,\mathbf{p}_B$ from the set of all  non-negative  four-vectors 
$\left\{\mathbf{p}_B \vert\, \mathbf{p}^T_B \,  G\, \mathbf{p}_B\geq 0, \, p_{B_0}> 0\, \right\}$
 to the set 
$\left\{\mathbf{q}_A=\Lambda\,\mathbf{p}_B\,\vert\, \mathbf{q}^T_A\,G\, \mathbf{q}^T_A\geq 0,\, q_{A_0}> 0\,\right\}$. 

Using the fact that every positive four-vector can always be expressed as a sum of neutral four-vectors~\cite{KNS,AVG1, AVG2}, we conveniently restrict ourselves to the maps  
\begin{itemize}
\item[(i)]   $\mathbf{p}_{n}\mapsto\mathbf{q}=\Lambda\,\mathbf{p}_{n}$
\item [(ii)] $\mathbf{p}_{n}\mapsto\mathbf{q}=\Lambda^T\,\mathbf{p}_{n}$
\end{itemize}  
 induced by the real matrix  $\Lambda$, on the set of all  neutral four-vectors 
$\{\mathbf{p}_{n}\vert \mathbf{p}_{n}^T\,G\, \mathbf{p}_{n}=0, p_{n0}> 0\}.$

Let us consider the set of all neutral four-vectors $\left\{\mathbf{p}_{n}=\left(1, x_1,x_2,x_3\right)^T,\, x_1^2+x_2^2+x_3^2=1\right\}$, with  $(x_1,x_2,x_3)$ representing the entire Bloch sphere (i.e., the unit sphere ${\mathcal S}^2\in \mathbbm{R}^3$). The type-I canonical form  $\Lambda^{{\rm I}_c}_{A}$ given in (\ref{diagAIc}) transforms $\mathbf{p}_{n}=\left(1, x_1,x_2,x_3\right)^T$ to a non-negative four-vector  $\mathbf{q}~=~ \Lambda^{{\rm I}_c}\, \mathbf{p}_{n}  ~=~(1,\, y_{1},y_{2},y_{3})$ where 
\begin{eqnarray}
y_{1}&=&\sqrt{\frac{\lambda_1}{\lambda_0}}\, x_1,\ \  
  y_{2}=\sqrt{\frac{\lambda_2}{\lambda_0}}   x_2   \nonumber \\  
   y_{3}&=&\pm\sqrt{\frac{\lambda_3}{\lambda_0}}\, x_3. 
\end{eqnarray}  
Evidently,  the transformed three-vector $\left(y_{1},y_{2},y_{3}\right)$ obeys the equation of a point on the surface of an ellipsoid  
\begin{equation}
\label{elIB}
\frac{y^2_{1}}{\xi^2_{1}} + \frac{y^2_{2}}{{\xi^2_2}} +\frac{y^3_{3}}{{\xi^2_3}}=1 
\end{equation}
where $\xi_i=\sqrt{\lambda_i/\lambda_0}, \, i=1,2,3$. Geometric intuition of the canonical form $\Lambda^{{\rm I}_c}_{A}$ is thus clear:  the map  $\Lambda^{{\rm I}_c}: \left(1, x_1,x_2,x_3\right)^T \mapsto \left(1, y_{1},y_{2},y_{3}\right)^T$   transforms the Bloch sphere to an ellipsoidal surface described by (\ref{elIB}). It may be recognized that the ellipsoidal surface described by (\ref{elIB}) geometrically represents the set of all steered Bloch vectors~\cite{verstraete2002,shi2011,shi2012,jevtic2014} of Alice's (Bob's) qubit  after Bob (Alice) performs projective measurements     
  on his (her) qubit (see (\ref{mapB2A}),(\ref{pos3}), (\ref{pos4})), given that the two-qubit state shared between them is in the canonical Bell-diagonal form (\ref{bds}), which is  achieved  by  SLOCC on $\rho_{AB}$. 

In Fig.~1 we have depicted the ellipsoid with lengths of  its semi-axes given by (see (\ref{elIB}))
$\left(\sqrt{\lambda_1/\lambda_0},\, \sqrt{\lambda_2/\lambda_0},\, \sqrt{\lambda_3/\lambda_0}\right)$. Here $\lambda_0\geq\lambda_1\geq\lambda_2\geq \lambda_3$ The ellipsoid is centered at the origin $(0,0,0)$. 
\begin{figure}[h]
	\label{typeI}
	\begin{center}
		\includegraphics*[width=3in,keepaspectratio]{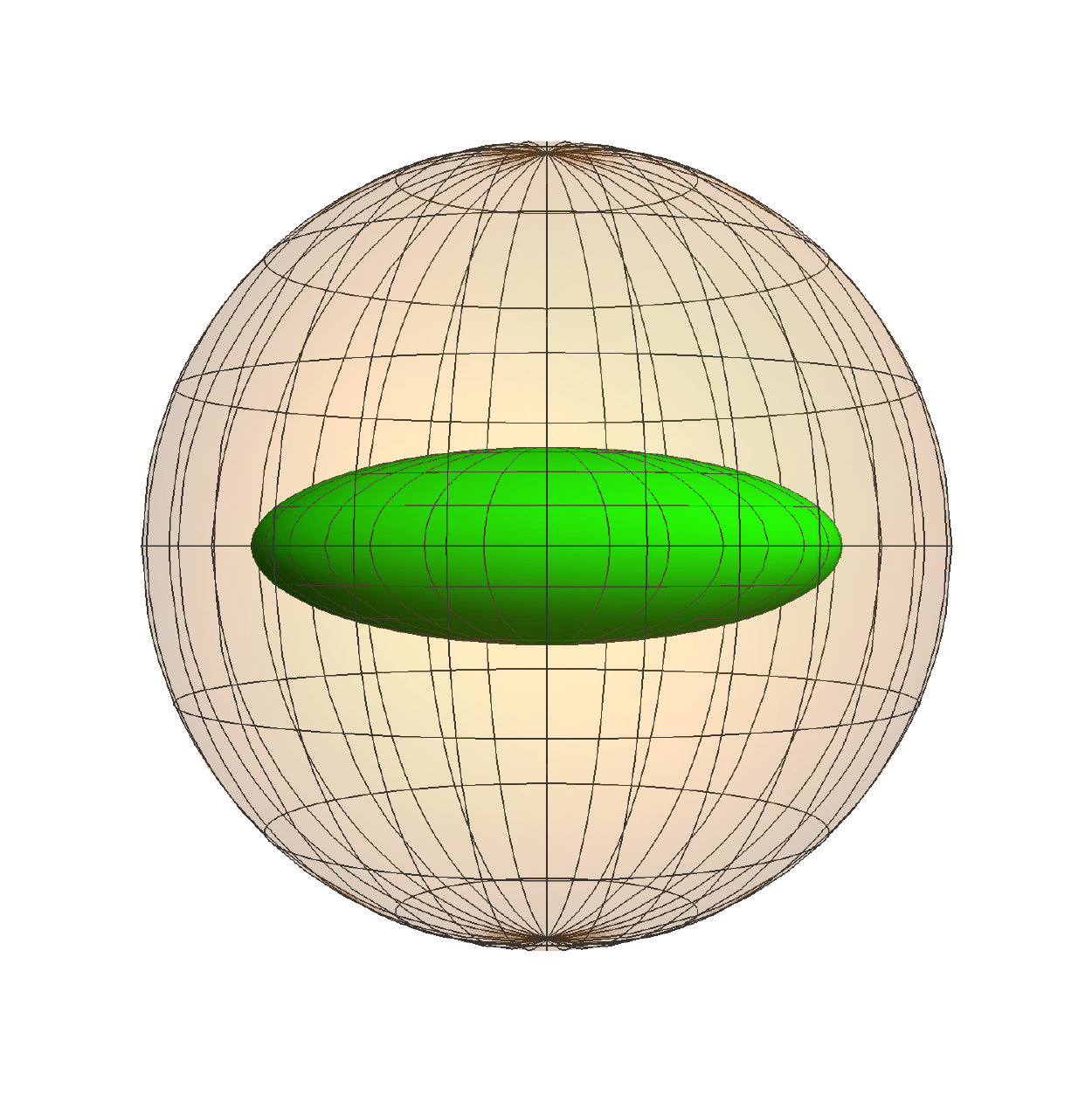}
		\caption{(Colour online) Ellipsoid  representing type-I canonical form $\Lambda^{{\rm I}_c}$ given by (\ref{diagAIc}).  Semi-axes lengths of this ellipsoid   (see (\ref{elIB})) are given by  $\left(\sqrt{\lambda_1/\lambda_0},\, \sqrt{\lambda_2/\lambda_0},\, \sqrt{\lambda_3/\lambda_0}\right)$,\ where $\lambda_0\geq\lambda_1\geq\lambda_2\geq \lambda_3$ denote eigenvalues of $G\Omega_A$, $G\Omega_B$ (see (\ref{goab})). The ellipsoid is centered at the origin $(0,0,0)$ and it provides geometric insight for  the set of all two-qubit states, which are on the SLOCC orbit of Bell-diagonal states (\ref{bds}).}  
	\end{center}
\end{figure}
      
Associated with the type-II canonical forms $\Lambda^{{\rm II}_c}_{A},\, \Lambda^{{\rm II}_c}_{B}$ (see (\ref{abfincanII})) one obtains 
\begin{eqnarray}
 \Lambda^{{\rm II}_c}_{A}\, (1,x_1,x_2,x_3)^T\, &=& (1,\,y_{A_1},\,y_{A_2},\, y_{A_3})^T. \nonumber  \\ 
\left(\Lambda^{{\rm II}_c}_{B}\,\right)^T\, (1,x_1,x_2,x_3)^T &=& (1,\,y_{B_1},\,y_{B_2},\, y_{B_3})^T \nonumber \\
\end{eqnarray}   
Here  $x_1^2+x_2^2+x_3^2=1$ represents the  Bloch sphere and 
\begin{eqnarray}
\label{IIellipsoid}
y_{A_1}&=& r_1\,x_1,\ y_{A_2}=-r_1\, x_2  \nonumber \\  
  y_{A_3}&=&(1-r_0)\, +\, r_0\, x_3,\ \ \ 0\leq r_1^2\leq r_0\leq 1  \\ 
y_{B_1}&=& s_1\,x_1,\ y_{B_2}=-s_1\, x_2 \nonumber \\ 
 \  y_{B_3}&=& (1-s_0)+s_0\, x_1,\ \ 0\leq s_1^2\leq s_0\leq 1 
\end{eqnarray}   
represent  the  set of all qubit states (Bloch vectors) that can be steered to by projective  measurements performed on another qubit of the two-qubit state 
$\rho_{AB}^{{\rm II}_c}$ of (\ref{abfincanII}), where $r_0, r_1,$  are specified by  (\ref{r0}), (\ref{plambda_a})  and $s_0,s_1$ are defined via (\ref{s0}), (\ref{plambda_b}), together with (\ref{p0}), (\ref{q0}).
\begin{figure}[h]
	\label{typeII}
	\begin{center}
		\includegraphics*[width=3in,keepaspectratio]{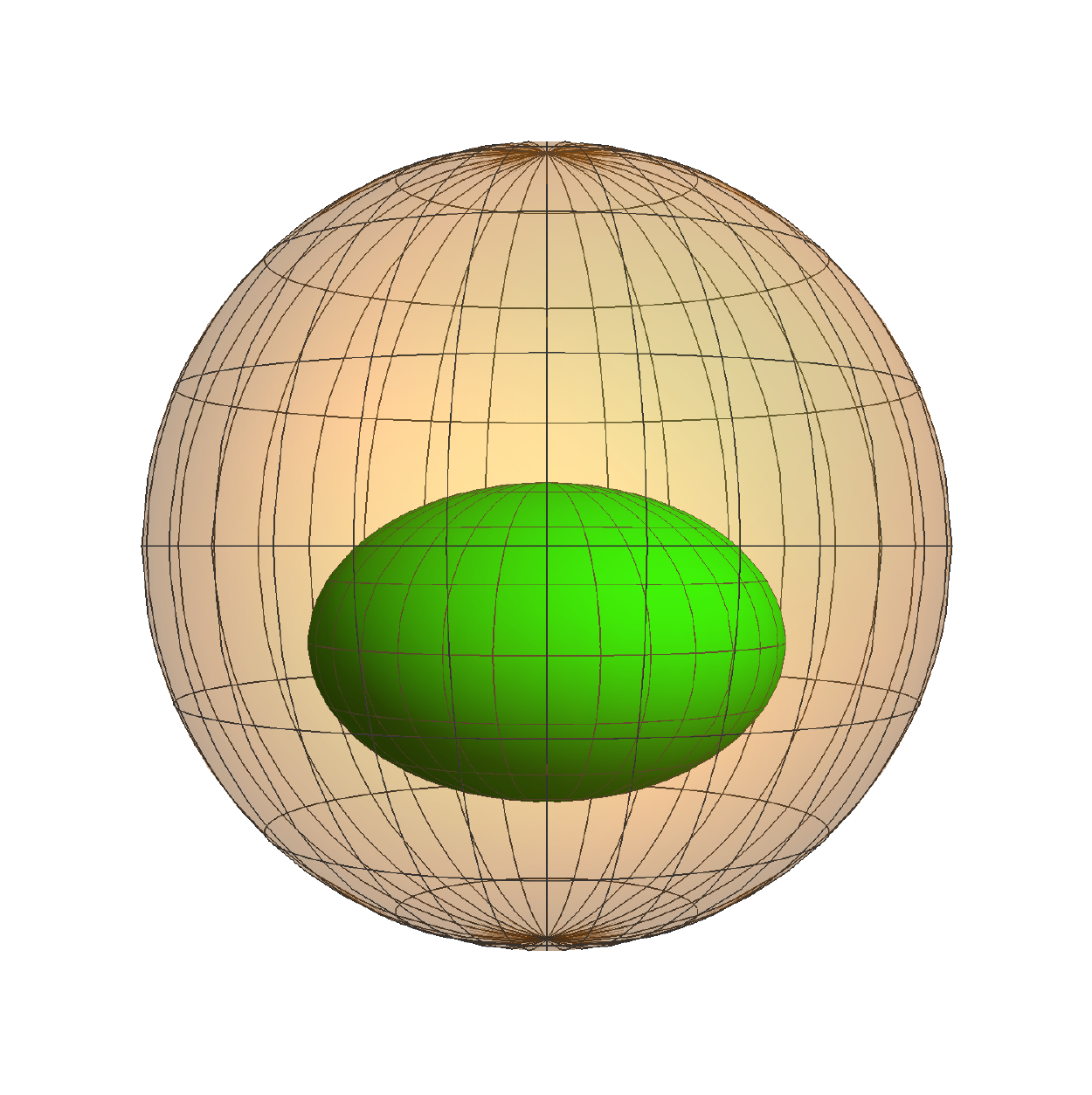}
		\caption{(Colour online) Steering spheroid (\ref{IIellipsoidA}) offering pictorial representation of type-II canonical form $\Lambda^{{\rm II}_c}_A$, which characterizes the   two-qubit state $\rho_{AB}^{{\rm II}_c}$  (see (\ref{abfincanII})). The spheroid is centered at $(0,0,1-r_0)$ and has semi-axes lengths $(r_1,r_1,r_0)$,\   $0\leq r_1^2\leq r_0\leq 1$.}
	\end{center}
\end{figure}
From (\ref{IIellipsoid}) it is seen that $\left(y_{A_1},y_{A_2}, y_{A_3}\right)$ and $\left(y_{B_1},y_{B_2}, y_{B_3}\right)$ satisfy the equations 
\begin{eqnarray}
\label{IIellipsoidA}
\frac{y^2_{A_1}+y^2_{A_2}}{r_1^2}+\frac{\left(y_{A_3}-(1-r_0)\right)^2}{r_0^2}&=&1  \nonumber \\
\label{IIellipsoidB}
\frac{y^2_{B_1}+y^2_{B_2}}{s_1^2}+\frac{\left(y_{B_3}-(1-s_0)\right)^2}{s_0^2}&=&1  
\end{eqnarray}   
which represent surfaces of  spheroids  centered respectively at $(0,0, 1-r_0)$, $(0,0, 1-s_0)$.   The spheroidal surfaces (\ref{IIellipsoidA}) provide geometric visualization of the collection of all Bloch vectors of one of the qubits after projective measurements are performed on the other qubit     ~\cite{verstraete2002,shi2011,shi2012,jevtic2014}, when the two-qubit state $\rho_{AB}$ is SLOCC equivalent to the type-II canonical density matrix $\rho^{{\rm II}_c}_{AB}$ of (\ref{abfincanII}). 
In Fig.~2  steering spheroid representing type-II  states $\rho_{AB}^{{\rm II}_c}$ of (\ref{abfincanII})  is shown.  

\section{Summary} 

In this paper we have presented a complete analysis to obtain two different types of SLOCC canonical forms and the associated geometric visualization of two-qubit states -- which happen to be the simplest  composite systems. Using the established result that the action of SLOCC on a two-qubit state  $\rho_{AB}=\frac{1}{4}\,\sum_{\mu,\nu=0}^3\, \Lambda_{\mu,\nu}\,\sigma_\mu\otimes\sigma_\nu$ manifests itself in terms of Lorentz transformation on its $4\times 4$ real matrix parametrization $\Lambda$,  two  different  types of canonical forms had been obtained previously by Verstraete et. al.~\cite{verstraete2001,verstraete2002}. 
However, the approach employed by  Ref.~\cite{verstraete2001,verstraete2002} to arrive at the  SLOCC canonical forms involved highly technical results  
on  matrix decompositions in spaces with indefinite metric. 

Based on a different approach, insprired by the techniques developed in classical polarization optics by some of us~\cite{AVG1,AVG2}, we have given here a simple procedure to explicitly evaluate two different types of SLOCC  canonical forms of the real matrix $\Lambda$ and the associated two-qubit density matrix. Equivalence between the canonical forms obtained via  our approach with the ones  obtained in Ref.~\cite{verstraete2001} has also been established here. Finally, our approach leads to an elegant geometric representation aiding visualization of two different types of canonical forms associated with the entire family of two-qubit states on the respective SLOCC orbits.  We believe that our comprehensive analysis  offers new insights in the study of  SLOCC canonical forms of higher dimensional and multipartite composite systems too.

\section*{Acknowledgements} Sudha and ARU are supported by the Department of Science and Technology(DST), India through Project No. DST/ICPS/QUST/Theme-2/2019 (Proposal Ref. No. 107). HSK acknowledges the support of NCN through grant SHENG (2018/30/Q/ST2/00625).
 This work was partly done when HSK was at The Institute of Mathematical Sciences, Chennai, India and completed when the author moved to  ICTQT, Gdansk, Poland.  KSA is supported by the University Grants Commission (UGC), India BSR-RFSMSKA scheme. 

\appendix
\section*{APPENDIX A} 
\renewcommand{\theequation}{A\arabic{equation}}
\setcounter{equation}{0}
For the sake of completeness we give a brief outline covering essential elements of the proof of the theorem stated in Sec.~3. For a detailed proof, with all its nuances addressed, see Ref.~\cite{AVG1}. 

To begin with, note that the real matrix $\Lambda$ parametrizing a two qubit density matrix  induces a linear trasnformation $\Lambda: \mathbf{p}\mapsto \mathbf{q}=\Lambda\,\mathbf{p}$ from the set of all  non-negative  four-vectors 
\be
\label{aa0}
\left\{\mathbf{p} \vert\, \mathbf{p}^T \,  G\, \mathbf{p}\geq 0, \, p_{0}> 0\, \right\}
\ee
to another identical set  
\be
\label{aa0'}
\left\{\mathbf{q}=\Lambda\,\mathbf{p}\,\vert\, \mathbf{q}^T\,G\, \mathbf{q}\geq 0,\, q_{0}> 0\,\right\}.
\ee 
Since every positive four-vector can always be expressed as a sum of neutral four-vectors~\cite{AVG1}, we restrict ourselves to the set 
$$\left\{\mathbf{p}_{n}=\left(1, x\right)^T,\, x^T x=x_1^2+x_2^2+x_3^2=1; \mathbf{p}_{n}^T\,G\,\mathbf{p}_{n}=0\right\}$$  without any loss of generality. We then express the non-negativity condition (\ref{aa0'}) as  
\begin{eqnarray} 
\label{aa1}
\left\{ \mathbf{q}=\Lambda\,\mathbf{p}_n,\  \mathbf{p}_{n}^T\,G\,\mathbf{p}_{n}=0 \Rightarrow \mathbf{q}^T\,G\, \mathbf{q}=\mathbf{p}_n^T\,\Omega\, \mathbf{p}_n\geq 0 \right\}, \nonumber \\   
\end{eqnarray} 
where 
\be
\Omega=\Lambda^T\,G\,\Lambda= \Omega^T
\ee
is a $4\times 4$ real symmetric  matrix.

Let us express $\Omega$ as a $1\oplus 3$ block matrix: 
\be
\label{omg}
\Omega=\ba{cc} n_0 & \tilde{n}^T \\  \tilde{n} & A  \ea 
\ee
with $n_0>0$, $\tilde{n}=(\tilde{n}_1,\tilde{n}_2,\tilde{n}_3)^T$ a 3 componental  column  and  $A^T=A$ is a $3\times 3$ real symmetric matrix. 

With the help of a Lorentz transformation  $L=1\oplus R$, where  $R\in$\,SO(3) denotes a three dimensional rotation matrix, one can diagonalize the $3\times 3$ real symmetric matrix $A$  (see (\ref{omg})) i.e.,  
$R^T\,A\,R=A_0=\mbox{diag}\left( \alpha_1,\,\alpha_2,\, \alpha_3\right)$. 
We thus obtain 
\begin{eqnarray*}
\Omega_0= L^T\,\Omega\,L&=&\ba{cccc} n_0 &  n_1 & n_2 & n_3  \\ 
n_1 & \alpha_1  & 0 & 0 \\ 
n_2 & 0 &  \alpha_1  & 0  \\
n_3 & 0 & 0 & \alpha_3   \ea \\ 
\end{eqnarray*}
where $(n_1,\,n_2,\,n_3)^T=n=R\, \tilde{n}.$

Let us denote  
$${\mathbf{p}_n}^T\,\Omega_0\,{\mathbf{p}_n}=D(\Omega_0;\, x).$$ 
The non-negativity condition ${\mathbf{p}_n}^T\,\Omega_0\,{\mathbf{p}_n}=D(\Omega_0;\, x)\geq 0$ assumes the form  
\be 
\label{ds1}
D(\Omega_0;\,x)=n_0+2\, x^T\,n + x^T A_0\ n\geq 0 \ \ \forall \ \ x^T\,x=1.
\ee 
Note that the condition (\ref{ds1})  is ensured  for all $x^T x=1$, if the absolute minimum $D_{\rm min}$ of the function  $D(\Omega_0;\,x)$, or equivalently, 
all the  {\emph {critical values}} $D_a$ of $D(\Omega_0;\,x)$ are    non-negative. The method of Lagrange multipliers to evaluate the critical values  of the function $D(\Omega_0;\,x)$, subject to the constraint $ x^T\,x=1$,  leads to an auxiliary function       
\be
\label{k1}
K(\Omega_0;\,x)=D(\Omega_0;\, x)+\lambda\, (x^T\,x-1)  
\ee 
where $\lambda$ denotes the Lagrange multiplier. Critical values $D_a$ of the function $D(\Omega_0;\,x)$ can then be obtained by solving    
\begin{eqnarray}
\label{lag}
\left.\frac{\partial K(\Omega_0;\,x)}{\partial \lambda}\right\vert_{\lambda_a,x_{a,i}}&=& 0 \nonumber \\ 
\left.\frac{\partial K(\Omega_0;\,x)}{\partial x_i}\right\vert_{\lambda_a,x_{a,i}}&=& 0, \ \ \ i=1,\,2,\,3.  
\end{eqnarray}
The equations determining  $\lambda_a$, $x_{a,i}=(x_{a,1},x_{a,2},x_{a,3})$ can then be expressed as 
\be
\label{decoup}
(A_0+\lambda_a\,\mathbbm{1}_3)\, x_a=- n,\ \    x_{a}^T\, x_a=1, 
\ee
where $\mathbbm{1}_3$ denotes $3\times 3$ identity matrix. 
Solutions of (\ref{decoup}) in turn determine the critical values $D(\Omega_0;\,\lambda_a,\,x_a)=D_a$ of the function $D(\Omega_0;\, x)$.

Substituting $A_0=\mbox{diag}\left( \alpha_1,\,\alpha_2,\, \alpha_3\right)$ in (\ref{decoup}) and simplifying, we obtain 
\be
\label{xc}
x_{a,i}=\frac{-n_i}{(\alpha_i+\lambda_a)},\  i=1,\,2,\,3.
\ee  
Furthermore, the normalization condition $x_{a}^T\, x_a=1$ leads to 
\be
\label{lamc}
\sum_{i=1}^3\, \frac{n_i^2}{(\alpha_i+\lambda_a)^2}=1.
\ee   
The critical values of $D(\Omega_0;\, x)$ are then given by   
\be  
\label{d2}
D_a=n_0-\lambda_a-\sum_{i=1}^3\, \frac{n_i^2}{\lambda_a+\alpha_i}.  
\ee 

We focus on identifying the implications of the conditions $D_a\geq 0,\, a=1,2,\ldots $ on the eigenvalues and eigenvectors of the $4\times 4$  matrix $G\,\Omega$, which are termed as G-eigenvalues and G-eignevectors of the real symmetric matrix $\Omega$.  To this end, we study  the behavior of the function  
\be
\label{hfunction}
h(\lambda)=n_0-\lambda-\sum_{i=1}^3\, \frac{n_i^2}{\lambda+\alpha_i}, 
\ee 
obtained by replacing $\lambda_a$ by a continuous real variable $\lambda$ in (\ref{d2})).

We list some of the important properties of the function  $h(\lambda)$, which follow from its definition (\ref{hfunction}):   
\begin{itemize}
	\item[(a)] The function $h(\lambda)$ is differentiable everywhere on the $\lambda$-axis except for a finite number of discontinuities at $\lambda~=~-~\alpha_i,\, i=1,2,3 $, whenever the corresponding $n_i~\neq~0$.  
	\item[(b)] As $h(\lambda)$ changes sign across a discontinuity, it is positive to the immediate left and is negative to the immediate  right of a discontinuity. This implies that there must be an odd number of real zeros of the function  $h(\lambda)$ in  
	between any two consecutive discontinuities.
	\item[(c)] In the limit  $\lambda\rightarrow \infty$ it is seen that $h(\lambda)\rightarrow -\infty$ and as $\lambda\rightarrow -\infty$ one finds  $h(\lambda)\rightarrow \infty$.   
	This observation along with the behaviour of $h(\lambda)$ near a discontinuity leads to the conclusion that there must be  even number of zeroes in the interval $(\alpha_{\rm max},\,\infty)$. 
	\item[(d)] Since the largest zero $\lambda_{\rm max}$ occurs in the interval $(\alpha_{\rm max},\,\infty)$, the slope of $h(\lambda)$ at $\lambda_{\rm max}$ is either negative or zero. In fact, when $h'(\lambda_{\rm max})=0$, both  the zero and the critical value  occur simultaneously at    $\lambda_{\rm max}$. (Here $h'(\lambda)$ denotes differentiation of $h(\lambda)$ with respect to the variable $\lambda$).   
	\item[(e)] The function $h(\lambda)$ must have at least $k+1$ real zeros where $k\leq 3$ denotes the number of discontinuities.   
	\item[(f)] Depending on the number of non-zero values of $n_1,n_2,n_3$ and based on the degeneracies  $\alpha_1$, $\alpha_2$, $\alpha_3$, there are  $20$ possible situations, each  with different number of discontinuities, zeroes and the critical values of 
	$h(\lambda)$:  (i) none of $n_1,n_2,n_3$ are zero;  (ii) one of  $n_1,n_2,n_3$ is zero; 
	(iii) two of  $n_1,n_2,n_3$ are zero; 
	(iv)  $n_1=n_2=n_3=0$.  Each of these 4 cases fall under 5 different subclasses corresponding to the degeneracies of $\alpha_1,\alpha_2,\alpha_3$:   non-degenerate i.e.,  (A)~$\alpha_1\neq\alpha_2\neq \alpha_3$,  two-fold degenerate i.e,  \break (B1)~$\alpha_1=\alpha_2\equiv\alpha\neq \alpha_3$,  (B2)~$\alpha_1\neq\alpha_2= \alpha_3\equiv\alpha$, 
	(B3)~$\alpha_1=\alpha_3\equiv\alpha\neq \alpha_2$, and  fully degenerate i.e., (C)~$\alpha_1=\alpha_2=\alpha_3=\alpha$. 
	
	Associated with these $4\times 5=20$ distinct possibilities one may list the number of discontinuities, zeros, local maxima and local minima of $h(\lambda)$. As mentioned already  there are $k+1$ real zeros  associated with $k$ discontinuities of $h(\lambda)$. 
	For instance,  if there are $k=3$ distinct discontinuities (realized when $\alpha_1\neq \alpha_2\neq \alpha_3$ and $n_1, n_2, n_3\neq 0$), it can be seen that at least {\em two} zeros exist. Furthermore, in the region ($\alpha_{\rm max},\,\infty$) one should find at least {\em two} zeroes. In other words, at least {\em four} real zeros exist for the function $h(\lambda)$ when there are {\rm three} distinct discontinuities. When there are {\em two} distinct discontinuities ($k=2$), at least {\em one} zero of the function $h(\lambda)$ occurs between them;  in the region ($\alpha_{\rm max},\,\infty$)  {\em two} zeroes (either distinct or doubly repeated) exist. Thus,  $1+2=3$ real zeroes exist for $h(\lambda)$ when $k=2$.      
\end{itemize} 

Interestingly, the function $h(\lambda)$ can be expressed in terms of the characteristic polynomial $\phi(\lambda)~=~\mbox{det}~(\Omega~-~\lambda\,G~)$   
of the real symmetric matrix $\Omega$ and $\psi(\lambda)=\prod_{i=1}^{3}\,(\alpha_i+\lambda)=\mbox{det}\,(A_0+\lambda\,\mathbbm{1}_3)$ as:   
\begin{eqnarray}
\label{hfndet} 
h(\lambda)&=& \frac{\phi(\lambda)}{\psi(\lambda)} = \frac{\mbox{det}(\Omega_0-\lambda\,G)}{\mbox{det}\,(A_0+\lambda\,\mathbbm{1}_3)}\nonumber \\
&=& \frac{\mbox{det}(\Omega-\lambda\,G)}{\mbox{det}\,(A+\lambda\,\mathbbm{1}_3)}
\end{eqnarray} 
Furthermore, it is found convenient to express the characteristic polynomial $\phi(\lambda)$  as  
\begin{eqnarray*}
	\phi(\lambda)&=&\psi(\lambda)h(\lambda) \nonumber \\ 
	&=& \phi_1(\lambda)\,g(\lambda) h(\lambda) 
\end{eqnarray*}
in terms of some simple polynomials  $\phi_1(\lambda)$, $g(\lambda)$ with real roots, chosen such that the  roots of $\phi_1(\lambda)$ may be readily identified and $\phi_1(\lambda)$, $g(\lambda)$ are finite at every real zero of the function  $h(\lambda)$.  

Examining the characteristic equation $\phi(\lambda)=0$  and based on explicit evaluations of $\phi_1(\lambda)$, $g(\lambda)$ and $h(\lambda)$ in each of the 20 cases one arrives at the following conclusions~\cite{AVG1}: 
\begin{enumerate} 
	\item Every real zero of $h(\lambda)$ is a G-eigenvalue of $\Omega$. 
	\item If $r$ denotes the number of (real) roots of $\phi_1(\lambda)$ and $k$ denotes the number of discontinuities of $h(\lambda)$  then it is identified that  $r+k+1=4$ in all the 20 cases, thus proving that $\phi(\lambda)$  has four real roots $\lambda_\mu,\mu=0,1,2,3$. This proves that the G-eigenvalues of $\Omega$ are real.
	\item  If $\mathbf{x}$ denotes the G-eigenvector of $\Omega$ belonging to G-eigenvalue $\lambda$, it can be seen that
	\be 
	\label{x0}
	{\mathbf x}^TGX=-h'(\lambda).
	\ee 
	Let us denote the largest G-eigenvalue of $\Omega$ by $\lambda_0$. As stated already (see property (d) of the function $h(\lambda)$)  $h'(\lambda_0)$ must be either negative or zero. Thus, from (\ref{x0}) it is clear that  the G-eigenvector ${\mathbf x}_0$ belonging to the largest G-eigenvalue $\lambda_0$ obeys ${\mathbf x}^TG{\mathbf x}\geq 0$ implying that it is either positive or neutral.  
	
	It also follows that the largest G-eigenvalue $\lambda_0$  is doubly degenerate when $h'(\lambda_0)=0$. In other words,  ${\mathbf x}_0$ corresponding to a largest doubly degenerate eigenvalue $\lambda_0$ is a neutral four-vector. 
	\item The G-eigenvectors $\mathbf{x}_r$  corresponding to the  G-eigenvalues $\lambda_r~<~\lambda_0$ of $\Omega$ are negative i.e., $\mathbf{x}_r^T\,G\,\mathbf{x}_r<0$.  This follows essentially from the observation  that $h'(\lambda_r)~=~-~\mathbf{x}_r^T\,G\,\mathbf{x}_r$   (see (\ref{x0}) is positive when $\lambda_r<\lambda_0$. 
	\item An explicit analysis of the G-eigenspace of $\lambda_0$ in each of the 20 different cases proves that 
	the real symmetric $4\times 4$ matrix $\Omega$, obeying the condition $\mathbf{p}_n^T\,\Omega\mathbf{p}_n\geq 0$,  possesses either  
	(i) a positive G-eigenvector belonging to the largest G-eigenvalue $\lambda_0$ and three negative G-eigenvectors or 
	(ii) a neutral G-eigenvector belonging to {\emph {at least}} doubly degenerate G-eigenvalue $\lambda_0$ and two negative 
	G-eigenvectors. 
	\item  A tetrad consisting of one positive and three negative G-eigenvectors constitute the columns of a Lorentz matrix  which ensures the transformation  $\Omega\rightarrow \Omega^c=L\Omega^{{\rm I}_c} L^T$ to a diagonal canonical form $\Omega^{{\rm I}_c}$. Based on a triad  consisting of one neutral and two negative G-eigenvectors it is possible to construct a Lorentz matrix (see Appendix B where explicit construction of Lorentz matrix in this case is given) such that  transformation  $\Omega\rightarrow \Omega^c=L\Omega^{{\rm II}_c} L^T$ resulting in a non-diagonal canonical form $\Omega^{{\rm II}_c}$ can be obtained (when the largest eigenvalue $\lambda_0$  of $\Omega$ is doubly degenerate and the corresponding G-eigenvector is neutral). 
	\item Using the explicit forms of the diagonal and non-diagonal canonical forms $\Omega^{{\rm I}_c}$ and $\Omega^{{\rm II}_c}$ of $\Omega$  it can be explicitly verified  that the G-eigenvalues of $\Omega$ are {\em non-negative}. 
\end{enumerate} 

\section*{APPENDIX B} 
\renewcommand{\theequation}{B\arabic{equation}}
\setcounter{equation}{0}
In this appendix we discuss  explicit construction of a Lorentz matrix $L$ belonging to OPLG in terms of a set of  G-orthogonal  four-vectors~\cite{KNS,AVG1,AVG2}.  
\begin{enumerate}[label=(\roman*)]
	\item  Consider a positive   four-vector $\mathbf{x}_0$ with its zeroth component $x_{00}> 0$,  and three other negative four-vectors $\mathbf{x}_i,\ i=1,2,3,$ obeying Minkowski G-orthogonality conditions i.e.,  
	\begin{equation}
	\label{tsss}
	\mathbf{x}_\mu^T\, G\, \mathbf{x}_\nu=G_{\mu\,\nu}, \mu,\nu=0,1,2,3, 
	\end{equation} 
	where $G_{\mu\,\nu}$ denotes elements of the Minkowski matrix $G$.
	The  set $\{\mathbf{x}_\mu, \mu=0,1,2,3\}$ of four-vectors           
	obeying (\ref{tsss}) forms  an {\em $G$-orthogonal tetrad} in $\cal{M}.$    
	
	It is readily seen that a real $4\times 4$ matrix $L~=~\left(\begin{array}{llll} \mathbf{x}_0 & \mathbf{x}_1 & \mathbf{x}_2 & \mathbf{x}_3 \end{array} \right)$, with its columns forming a $G$-orthogonal set satisfies  $L^T\, G\, L=\, G$ with $(L)_{00}=x_{00}\geq 0$, and hence $L$ is a Lorentz matrix belonging to OPLG.
	
	\item A set $\{\mathbf{y}_0, \mathbf{y}_1, \mathbf{y}_2\}$  of four-vectors, consisting of a neutral vector $\mathbf{y}_0$ and two negative vectors $\mathbf{y}_1$, $\mathbf{y}_2$ obeying the property 
	\begin{eqnarray}
	\label{nss}
	\mathbf{y}_0^T\, G\, \mathbf{y}_0&=&0,\ \ \mathbf{y}_0^T\, G\, \mathbf{y}_i=0, \nonumber \\
	\mathbf{y}_i^T\, G\, \mathbf{y}_j&=&-\delta_{ij},\ \ i,j=1,2 \ \nonumber 
	\end{eqnarray} 
	forms a {\em G-orthogonal triad}. The neutral vector $\mathbf{y}_0$ is a self-orthogonal vector as its Minkowski norm   $\mathbf{y}^T_0\, G\, \mathbf{y}_0$ is zero. 
	
	Given the G-orthogonal triad $\{\mathbf{y}_0, \widetilde{\mathbf{y}}_1, \widetilde{\mathbf{y}}_1\}$,  consisting of a neutral  vector $\mathbf{y}_0$, it is possible  to construct a  tetrad  $\{\widetilde{\mathbf{y}}_0, \widetilde{\mathbf{y}}_1, \widetilde{\mathbf{y}}_2, \widetilde{\mathbf{y}}_3\, \}$ of four-vectors obeying the G-orthonormality conditions 
	$\widetilde{\mathbf{y}}_\mu^T\, G\, \widetilde{\mathbf{y}}_\nu=G_{\mu\, \nu},\ \mu,\nu=0,1,2,3$. To this end, we construct a four-vector $\mathbf{y}_3$ such that 
	\begin{eqnarray}
	\mathbf{y}_3^T\, G\, \mathbf{y}_0&\neq& 0,  \ \ 
	\mathbf{y}_3^T\, G\, \widetilde{\mathbf{y}}_i=0,\  i=2,3,
	\end{eqnarray}
	and define two four-vectors $\widetilde{\mathbf{y}}_0$ and $\widetilde{\mathbf{y}}_3$ as follows~\cite{AVG1, AVG2}: 
	\begin{eqnarray}
	\label{type2L}
	\widetilde{\mathbf{y}}_0&=&\mathbf{y}_3\, +\, \tau_y\,\mathbf{y}_0,\ \  y_{00}\geq 0  \nonumber \\ 
	\widetilde{\mathbf{y}}_3&=&\mathbf{y}_3\, -\, \kappa_y\,\mathbf{y}_0,
	\end{eqnarray}
	where the real parameters $\tau_y,\, \kappa_y$ are given by            
	\begin{eqnarray} 
	\tau_y= \frac{1-\mathbf{y}_3^T\, G\, \mathbf{y}_3}{2\, \mathbf{y}_3^T\, G\, \mathbf{y}_0},\ \   
	\kappa_y= \frac{1+\mathbf{y}_3^T\, G\, \mathbf{y}_3}{2\, \mathbf{y}_3^T\, G\, \tilde{\mathbf{y}}_0}. 
	\end{eqnarray}
	By construction, the set $\{\widetilde{\mathbf{y}}_0,\, \widetilde{\mathbf{y}}_1,\, \widetilde{\mathbf{y}}_2,\, \widetilde{\mathbf{y}}_3\}$ of four-vectors forms a G-orthonormal tetrad consisting of {\em one} positive and {\em three} negative four-vectors.  By following the explicit procedure outlined above one can construct a Lorentz matrix $L_2~=~\left(\begin{array}{llll} \widetilde{\mathbf{y}}_0 & \widetilde{\mathbf{y}}_1 & \widetilde{\mathbf{y}_2} & \widetilde{\mathbf{y}}_3 \end{array} \right)$, starting from a G-orthogonal  triad $\{\mathbf{y}_0, \widetilde{\mathbf{y}}_2, \widetilde{\mathbf{y}}_3\}$, consisting of a neutral four-vector $\tilde{\mathbf{y}}_0$. 
\end{enumerate}

\appendix


\begin{thebibliography}{99}
\bibitem{note2} A  real $4\times 4$ matrix $M$, transforming the set  of all positive and neutral four-vectors $\mathbf{s}\in{\mathcal{M}}$  with positive zeroth component  $\left\{\mathbf{s}\left\vert s_0> 0\right.\right\}$ to an identical set $\left\{M\mathbf{s}\left\vert (M\mathbf{s})_0\, > 0\right. \right\}$ is called a Mueller matrix in classical polarization optics~\cite{AVG1,AVG2}.
\end{thebibliography}
\end{document}